\documentclass[12pt]{article}

\usepackage[title]{appendix}

\usepackage[dvips]{graphicx} 
\usepackage{latexsym}
\usepackage[authoryear]{natbib}
\let \chapter \section
\usepackage{enumerate}
\usepackage{xfrac}
\usepackage{lmodern}
\usepackage{epsfig}
\usepackage{graphics,graphicx,color}
\usepackage{amssymb,amsfonts,amsthm,amsmath}
\usepackage{multirow, booktabs}
\usepackage{mathrsfs}

\usepackage{soul}       
\usepackage{float}       

\usepackage{caption}        
\usepackage{subcaption}     
\usepackage[table,xcdraw]{xcolor}
\usepackage{tabularx}

\newcommand{\diag}{\operatorname{diag}}

\usepackage{amsmath,bbm,amssymb,mathrsfs}
\usepackage{amsthm,amscd,float}             
\usepackage{graphicx, multirow, graphics}
\usepackage{comment}
\usepackage{longtable, lscape}
\usepackage{rotating}
\usepackage{dcolumn}
\usepackage{natbib}
\usepackage{booktabs,caption,fixltx2e}
\usepackage[flushleft]{threeparttable}
\usepackage{caption}
\usepackage{algorithm}
\usepackage{algpseudocode}

\newcommand{\bm}[1]{\mbox{\boldmath$ #1 $\unboldmath}}

\def\tr{{\rm tr}}
\def\vec{{\rm vec}}
\def \diag{{\rm diag}}

\date{\today}
\title{\bf Bayesian Sparse Regression for Mixed Multi-Responses with Application to Runtime Metrics Prediction in Fog Manufacturing} 
\author{Xiaoyu Chen$^{1}$, Xiaoning Kang$^{2}$, Ran Jin$^{3}$ and Xinwei Deng$^{4}$\\
$^{1}$Department of Industrial Engineering, University of Louisville, USA.\\
$^{2}$International Business College and Institute of Supply Chain Analytics, \\
Dongbei University of Finance and Economics, China \\
$^{3}$Grado Department of Industrial and Systems Engineering, \\
Virginia Tech, USA \\
$^{4}$Department of Statistics, Virginia Tech, USA
}
\date{} 

\begin{document}
\maketitle

\begin{abstract}
Fog manufacturing can greatly enhance traditional manufacturing systems through distributed Fog computation units,
which are governed by predictive computational workload offloading methods under different Industrial Internet architectures.
It is known that the predictive offloading methods highly depend on accurate prediction and uncertainty quantification of runtime performance metrics, containing multivariate mixed-type responses (i.e., continuous, counting, binary).
In this work, we propose a Bayesian sparse regression for multivariate mixed responses to enhance the prediction of runtime performance metrics and to enable the statistical inferences.
The proposed method considers both group and individual variable selection to jointly model the mixed types of runtime performance metrics.
The conditional dependency among multiple responses is described by a graphical model using the precision matrix, where a spike-and-slab prior is used to enable the sparse estimation of the graph.
The proposed method not only achieves accurate prediction, but also makes the predictive model more interpretable with statistical inferences on model parameters and prediction in the Fog manufacturing.
A simulation study and a real case example in a Fog manufacturing are conducted to demonstrate the merits of the proposed model.
\end{abstract}


\newpage
\section{Introduction}
Fog computing (also referred as Edge computing) techniques have served as an important role in Industrial Internet of things (IIoT) for smart manufacturing systems.
It provides local and distributed computation capabilities.
The concept of Fog manufacturing is defined on integrating a Fog computing network with interconnected manufacturing processes, facilitates, and systems.
With local computation units (i.e., Fog units) close to the manufacturing processes, the Cloud-based centralized computation architecture can be evolved to a Cloud-Fog collaborative computation to provide higher responsiveness and significantly lower time latency (\citealt{wu2017fog,zhang2019fog}).
There is a trade-off between the local computing efficiency on a Fog unit and the global collaborative efficiency of the centralized Cloud.
Specifically, the speciality of Fog units can significantly speedup the local computations, but it can pose significant challenges for the Cloud to assign the computation tasks and supervise the heterogeneous Fog units.
Besides, fluctuated computation capability of the Fog units and intermittent communication conditions among the Fog units and the Cloud
make it even harder for the collaboration (\citealt{zhang2015offloading}).
Therefore, computation offloading methods have been widely investigated to enable efficient collaboration between the Fog units and the Cloud with the consideration of constraints on resources.
In Fog manufacturing, the runtime performance metrics are often multivariate with mixed types (\citealt{chen2018predictive}).
These metrics include the CPU utilization (i.e., continuous response), temperature of the CPU (i.e., continuous response),
the number of computation tasks executed within a certain time period (i.e., counting response),
and whether the memory utilization exceeds certain thresholds (i.e., binary response).
Prediction and uncertainty quantification of these metrics are essential to support the computation in the Fog manufacturing, advancing analytics and optimization for high responsiveness and reliability (\citealt{wu2017fog, zhang2019fog}).
Based on the runtime performance metrics of these Fog nodes, the Fog computing can dynamically assign computation tasks to different Fog nodes (\citealt{chen2018predictive}).
The manufacturing must provide responsive and reliable computation services by meeting all requirements in runtime performance metrics.
It is thus of great importance to accurately predict runtime performance metrics of Fog nodes and quantify the uncertainty of prediction in task assignment and offloading problems.


As the runtime performance metrics are multivariate with mixed types, a simple method is to model each individual metric separately.
Clearly, such an approach overlooks the dependency relationship among the metrics, resulting in inaccurate prediction associated with high uncertainty.
For example, as the increment in the executed number of computation tasks per minute (i.e., counting response), the CPU utilization and temperature (i.e., continuous responses) will increase.
Quantifying such dependency among mixed responses is expected to improve the prediction accuracy.
Moreover, by only providing point estimation of mixed responses, the model prediction may not be trustworthy for those with high prediction variance.
Therefore, it calls for a joint model for the mixed responses with uncertainty quantification.
Towards predictive offloading, the objective is to jointly fit the mixed runtime performance metrics with the capability of statistical inferences to quantify uncertainties of the predicted metrics in Fog manufacturing.

In this work,  we propose a Bayesian sparse multivariate regression for mixed responses (BS-MRMR) to achieve accurate model prediction and, more importantly, to obtain proper statistical inferences of the responses.
The use of Bayesian estimation naturally enables uncertainty quantification of model prediction.
Both group sparsity and individual sparsity are imposed on regression coefficients via proper spike-and-slab priors.
The group structures often occur in the runtime performance metrics prediction problem when the metrics at the next time instance are regressed on two groups of predictors: the features extracted from the current and previous metrics (i.e., Group 1) and the covariates of the computation tasks (i.e., Group 2).
On the other hand, not all predictors are important within each group. Hence the individual sparsity is also induced for better estimation of model coefficients.
Moreover, the proposed method considers the conditional dependency among multiple responses by a graphical model using the precision matrix,
where a spike-and-slab prior is used to enable the sparse estimation of the graph.
A Gibbs sampling scheme is then developed to efficiently conduct model estimation and inferences for the proposed BS-MRMR method.
The proposed BS-MRMR model not only achieves accurate prediction, but also makes the predictive model more interpretable in the Fog manufacturing.
Note that one can consider a two-step Bayesian method to model the multivariate mixed responses \cite{bradley2022joint},
where the first step transforms the multivariate mixed-responses to continuous responses, and the second step models the transformed responses.
However, the obtained model coefficients are less interpretable since the transformation typically change the scale of the original responses.

Different from the recent work of \cite{kang2020multivariate} on a penalized regression for multivariate mixed responses,
the proposed BS-MRMR is a Bayesian approach with the following key novelty.
First, \cite{kang2020multivariate} only imposes individual sparsity while the BS-MRMR model takes into account of both the group and individual sparsity.
Second, the model introduced by \cite{kang2020multivariate} cannot  provide statistical inferences, such as prediction intervals for the responses due to their complicated parameter estimation procedure.
In contrast, the proposed BS-MRMR model is able to quantify the uncertainty of the estimated parameters and predicted responses within the Bayesian framework.
It provides a comprehensive information of prediction and uncertainty quantification to support the predictive offloading in Fog manufacturing.
Third, a careful investigation of the posterior distribution makes the computation of the Gibbs sampling efficient for model estimation and inference.


The remainder of this work is organized as follows. The proposed BS-MRMR model and the Gibbs sampling scheme are detailed in Section~\ref{sec:model}.
A simulation study is conducted to validate the BS-MRMR model in Section~\ref{sec:simu}. Section~\ref{sec:case} describes a real case study in Fog manufacturing.
Section~\ref{sec:conclusion} concludes this work with some discussions of future directions.


\section{Literature Review}

The joint modeling of mixed responses has attracted great attention in the literature.
Various existing studies focused on the bivariate responses.
For example, \cite{fitzmaurice1995regression} considered a bivariate linear regression model with a continuous and a binary response via joint likelihood estimation.
\cite{yang2007regression} proposed to jointly fit a continuous and a counting response, and evaluated the correlation between the bivariate responses varying over time through a likelihood ratio test.
These methods usually factorize the joint distribution of two responses as the product of a marginal and a conditional distribution (\citealt{cox1992response}),
which cannot be easily generalized for multivariate mixed responses in real applications such as Fog manufacturing.
Another direction of handling the continuous and discrete variables is to consider the underlying latent variables for the discrete responses, and then assume a multivariate normal distribution for such latent variables together with other continuous responses.
For example, \cite{regan1999likelihood} introduced a latent variable with a probit link function for a binary response and jointly modeled the continuous response and the latent variable via a bivariate normal distribution.
More related works include \cite{mcculloch2008joint, deng2015qq, wu2018sparse, kang2020multivariate}.
The advantage of introducing latent variables to characterize discrete responses lies mainly in the well-defined correlation measures among multivariate normal responses.
Therefore, the hidden association between mixed responses can be quantified by this correlation.
However, such models involving latent variables are often computationally expensive, especially when the number of predictor variables is large.

Modeling the mixed responses under the Bayesian framework is also studied in the literature.
For example, \cite{fahrmeir2007bayesian} fitted ordinal and normal responses via a Bayesian latent variable method,
where covariate effects on the latent variables were modeled through a semiparametric Gaussian regression model.
\cite{yeung2015bayesian} studied a dose-escalation procedure in clinical trials by a Bayesian approach,
where a logistic regression and a linear log-log relationship were used respectively to model the binary and continuous responses.
\cite{kang2018bayesian} proposed to fit the binary response conditioned on the continuous response, where proper priors were used for enhancing model interpretation.
However, few Bayesian works have been conducted for the multivariate mixed responses.
\cite{li2016bayesian} introduced a Bayesian conditional joint random-effects model for fitting longitudinal data with normal, binary and ordinal responses by using latent variables for each response.
More Bayesian methods can be found in \cite{dunson2000bayesian, zhou2006bayesian, stamey2013bayesian, hwang2014semiparametric, deyoreo2018bayesian}, among others.

There are few works of theoretical investigation on models of mixed responses under the framework of either conditional models or latent variables.
In the simple case where there are only one binary and one continuous responses,
\cite{kurum2016time} adopted latent variable to characterize the binary response and studied the asymptotic normality of their estimator.
Recently, \cite{kang2022generative} developed a generative model framework in which the continuous response is fitted based on the multivariate normal property, and the multi-class response is modeled and predicted via the linear discriminant analysis. They established the asymptotic properties of their estimator in terms of both the classification accuracy for the multi-class response and the prediction accuracy for the continuous response under some regularity conditions.

In addition, the proper regularization on model parameters is often used in the joint modeling of mixed responses for high-dimensional data to improve the model interpretation.
\cite{kang2020multivariate} proposed to fit data with multiple mixed responses by imposing $L_1$ penalties on the negative log-likelihood function and conducted the parameter estimation based on the EM algorithm.
Under the Bayesian framework, the spike-and-slab prior is commonly used for inducing the sparsity in regression models (\citealt{wagner2010bayesian}).
In our application of runtime performance metrics prediction problem, particularly, the group variable selection is necessary since predictor variables are naturally grouped by different components (e.g., CPU, RAM, etc.) of Fog units.
Hence the model coefficients may not be properly estimated across different subsets of samples if only the individual sparsity is imposed on the model.
In this regard, applying a group sparsity on predictor variables is important, especially when the multivariate responses are presented.

\section{The Proposed Bayesian Sparse MRMR} \label{sec:model}
For the proposed BS-MRMR model, we make the following assumptions.
First, assume the predictor variables are categorized into multiple groups and that a predictor is significant requires its variable group is significant.
Second, assume that the distributions of response variables are from the exponential family.
Third, assume that the hidden associations among mixed response variables can be represented in the precision matrix of latent Gaussian distributed variables.
Now we present the details of the proposed BS-MRMR model.

Suppose that the predictor variables are $\bm x = (X_{1}, \ldots, X_{p})^T$ and the multivariate mixed responses are
$\bm{Y} = (\bm{U}^T, \bm{Z}^T, \bm{W}^T)^T$.
Here $\bm{U} = (U^{(1)}, U^{(2)}, \ldots, U^{(l)})^{T}$ are the $l$-dimensional continuous responses, $\bm{Z} = (Z^{(1)}, Z^{(2)}, \ldots, Z^{(m)})^{T}$ are the $m$-dimensional counting responses,
and $\bm{W} = (W^{(1)}, W^{(2)}, \ldots, W^{(k)})^{T}$ are the $k$-dimensional binary responses.
To model the relationship between the predictor vector $\bm x$ and the response vector $\bm Y$,
we consider a generalized linear model (GLM) for each individual response under appropriate link functions,
while their link functions of the mean parameters form a multivariate linear model with respect to $\bm x$.
Specifically,
\begin{align}\label{eq:response}
	& U^{(j)} | \mu^{(j)}, {\sigma^{(j)}}  \sim N\left(\mu^{(j)}, {\sigma^{(j)}}^{2}\right), j = 1, \ldots, l, \nonumber \\
    & Z^{(j)} | \lambda^{(j)} \sim
\text{Poisson}\left(\lambda^{(j)}\right), j = 1, \ldots, m,  \nonumber \\
	& W^{(j)} | \gamma^{(j)}  \sim \text{Bernoulli}\left(\gamma^{(j)}\right), j = 1, \ldots, k,
\end{align}
and
\begin{align*}
\bm{\xi}  = (\mu^{(1)}, \ldots, \mu^{(l)}, \log\lambda^{(1)}, \ldots, \log\lambda^{(m)}, \log\frac{\gamma^{(1)}}{1 - \gamma^{(1)}}, \ldots, \log\frac{\gamma^{(k)}}{1 - \gamma^{(k)}})^T \sim \mathcal{N}_q(\bm{B}^T \bm{x}, \bm{\Omega}^{-1}),
\end{align*}
where $\bm B = (\beta_{ij})_{p \times q}$ is a $p \times q$ coefficient matrix with $q = l+m+k$.
Here $\mathcal{N}_q$ represents a $q$-dimensional multivariate normal distribution, and
$\bm \Omega$ is the precision matrix of the error term $\bm{\varepsilon}$ by defining $\bm{\xi} = \bm{B}^T \bm{x} + \bm{\varepsilon}$.
It is seen that the $\bm \xi$ is a latent vector connecting the multivariate mixed responses and predictor variables.
The implication of $\bm \Omega$ is to characterize the conditional dependency relationship among the multivariate mixed responses $\bm{U}$, $\bm{Z}$, and $\bm{W}$.
Denote the observed data as $(\bm x_{i}, \bm y_{i}), i=1, \ldots, n$.
Without loss of generality, we write $\bm \xi_i = \bm{B}^T \bm x_i + \bm{\varepsilon}_i$ with $\bm{\varepsilon}_i \sim \mathcal{N}_q(\bm{B}^{T} \bm{x}_i, \bm \Omega^{-1})$.
Hence the likelihood function can be expressed in a proper manner.
To conduct the parameter estimation from a Bayesian perspective, we need to specify the priors for parameter matrices $\bm B$ and $\bm \Omega$, respectively.

\subsection{Priors for Group and Individual Sparsity}\label{sec:estimation}
Since the latent vector $\bm \xi$ follows the normal distribution, one can consider a conjugate prior of normal distribution for parameters $\vec(\bm B)$, where $\vec(\cdot)$ is the vectorization operator.
However, such a prior does not encourage the sparsity on the coefficient matrix $\bm B$.
When fitting data with multivariate mixed responses, one would expect that only certain subgroups of predictor variables are related to the multivariate responses.
Therefore, we would like to impose an appropriate prior for matrix $\bm B$ to enable variable selection in the sense that only a few groups of predictor variables are selected and a few coefficients are nonzeros within each selected group. Here we assume that the grouping of predictor variables is known.
In particular, we propose to adopt a spike-and-slab prior \citep{liquet2017bayesian, ning2020bayesian} on the parameter matrix $\bm B$ for sparse estimation of parameters at both the group and individual levels.

Denote the data matrix by $\mathbb{X} = (\bm x_1, \bm x_2, \ldots, \bm x_n)^T$.
To facilitate the presentation, let the predictor vector $\bm x = (\bm X_1^T, \bm X_2^T, \ldots \bm X_G^T)^T$ to be composed of $G$ groups with $\bm X_g$ containing $p_g$ predictor variables for $g=1,2,\dots,G$.
Correspondingly, write $\mathbb{X} = (\mathbb{X}_1, \mathbb{X}_2, \ldots, \mathbb{X}_G)^{T}$ and the coefficient matrix $\bm B$ is partitioned as $\bm B = (\bm B_1^T, \bm B_2^T, \ldots, \bm B_G^T)^T$, where $\bm B_g$ is a $p_g \times q$ matrix for the $g$th group of predictor variables.
In order to enable variable selection for both group and individual levels,
we re-parameterize the coefficients matrix in each group as $\boldsymbol{B}_g=\bm V_g \Tilde{\boldsymbol{B}}_g$, where $\bm V_g=\diag{\{ \tau_{g,1},\dots,\tau_{g,p_g} \}}$ with $\tau_{g,j}\geq0$ for $j = 1, \ldots, p_{g}$.
The role of $\tau_{g, j}$ in diagonal matrix $\bm V_g$ is to control the sparsity of individual predictor variable within a group.
That is, $\tau_{g, j} = 0$ corresponds to the $j$th predictor variable in the $g$th group being excluded from the regression model.
Based on the above consideration, we employ the multivariate spike-and-slab prior for $\Tilde{\boldsymbol{B}}_g$ as:
\begin{align}
    \vec{(\Tilde{\boldsymbol{B}}^T_g | \bm{\Omega}, \pi_1}) &\sim
    (1-\pi_1) \mathcal{N}_{p_gq}(\boldsymbol{0},\boldsymbol{I}_{p_g} \otimes \bm{\Omega}^{-1}) +
    \pi_1\delta_0(\vec{(\Tilde{\boldsymbol{B}}^T_g)}),~~~g=1,\dots,G \label{prior:B} \\
    \tau_{g,j} | \pi_2, \sigma_{\tau}^2 &\sim (1-\pi_2) \mathcal{N}^+(0,\sigma_{\tau}^2) + \pi_2\delta_0(\tau_{g,j}),
    ~~~g=1,\dots,G; j=1,\dots, p_g \label{prior:tau}  \\
    \pi_1 &\sim \mbox{Beta}(a_1,a_2), \
    \pi_2 \sim \mbox{Beta}(a_3,a_4), \
    \sigma_{\tau}^2 \sim \mbox{IG}(1,d),  \nonumber
\end{align}
where $\bm I_{a}$ denotes the $a \times a$ identity matrix,
the notation $\otimes$ stands for the Kronecker product, the symbol $\delta_0(\cdot)$ is the Dirac measure that denotes the point mass at 0, the symbol $\mathcal{N}^+(0,\sigma_{\tau}^2)$ represents a normal distribution $\mathcal{N}(0,\sigma_{\tau}^2)$ truncated below at 0,
and IG$(a, b)$ is the inverse Gamma distribution with its density function $f(x) = b^a x^{-(a+1)} \exp(-b / x) / \Gamma(a)$.
The prior of $\Tilde{\boldsymbol{B}}$ in \eqref{prior:B} enables the variable selection at the group level,
with our prior belief of the entire group $\bm B_g$ excluding from the model by the probability parameter $\pi_1$.
Similarly, the prior of $\tau_{g,j}$ in \eqref{prior:tau} performs the variable selection at the individual level,
with our prior belief of the $j$th row of $\bm B_g$ excluding from the model by the probability parameter $\pi_2$.
Here we consider a Beta prior for $\pi_1$ to accommodate the potential domain-knowledge on the sparsity of the model.
When there is no pre-knowledge on which group of predictor variables are related with responses,
one could consider a simple uniform prior Unif(0,1) by setting $a_1 = a_2 = 1$.
In this work, we adopt the suggestion in \cite{scheipl2012spike} to use an informative Beta prior Beta(20, 40), which is suitable for the high-dimensional data.
Similarly, we adopt Beta(20, 40) as the prior of $\pi_2$ on the sparsity of individual predictor variable within each group.
In addition, the parameter $\sigma_{\tau}^2$ in the prior distribution of $\tau_{g,j}$ in \eqref{prior:tau} controls the shrinkage for the $j$th predictor variable in the $g$th group.
A large value of $\sigma_{\tau}^2$ may diffuse the coefficient for the corresponding predictor variable, and a small value may produce a biased estimated coefficient toward to zero.
We thus use a conjugate inverse gamma prior IG$(1,d)$ for $\sigma_{\tau}^2$ to determine its value from data,
and adopt the ``adaptive" idea for parameter $d$ by estimating it with the Monte Carlo EM algorithm, which is proposed by \cite{liquet2017bayesian}.
Specifically, in the $k$th EM iteration of estimating $d$, we update
$d^{(k)} = \mbox{E}^{-1}_{d^{(k-1)}}(1/\sigma_{\tau}^2 | \mbox{rest})$,
where the posterior expectation of $\sigma_{\tau}^2$ is replaced by the Monte Carlo sample average of $\sigma_{\tau}^2$ generated in the Gibbs samples based on $d^{(k-1)}$.

Next, we consider the prior of $\bm \Omega$ for inferring the conditional dependency relationship among responses.
The conventional Bayesian methods for imposing sparsity on $\bm \Omega$ are implemented by the priors over the space of positive definite matrices constrained by fixed zeros.
However, such priors often result in the daunting computational burdens for the large dimension of response variables.
To address this challenge, we adopt the prior of $\bm \Omega = (\omega_{ij})_{q \times q}$ in \eqref{prior:Omega}, which is a spike-and-slab prior similar as \cite{wang2015scaling} for efficient computation.
\begin{align} \label{prior:Omega}
\bm \Omega | \pi_3, \sigma_0^2, \sigma_1^2, \lambda &\sim
    \prod_{i<j} \left \{ (1-\pi_3) \mathcal{N} (\omega_{ij} ; 0, \sigma_0^2) +
    \pi_3 \mathcal{N} (\omega_{ij} ; 0, \sigma_1^2) \right \} \prod_{i} \mbox{e} (\omega_{ii} ; \frac{\lambda}{2}) \bm I ( \bm \Omega \in S^{+} )  \\
    \pi_3 &\sim \mbox{Beta}(a_5,a_6), \nonumber
\end{align}
where $\bm I(\cdot)$ is the indicator function, $S^{+}$ stands for the cone of symmetric positive definite matrices, e($\cdot$) denotes the exponential distribution,
and $\mathcal{N} (x; a, b)$ represents the density function of $\mathcal{N} (a, b)$ evaluated at point $x$.
The term $(1-\pi_3) \mathcal{N} (\omega_{ij} ; 0, \sigma_0^2) + \pi_3 \mathcal{N} (\omega_{ij} ; 0, \sigma_1^2)$ controls the sparsity on the off-diagonal elements $\omega_{ij}$, and the term e$(\omega_{ii} ; \lambda/2)$ shrinkages the diagonal elements $\omega_{ii}$.
The prior of $\bm \Omega$ in \eqref{prior:Omega} is computationally efficient since it can facilitate a fast block Gibbs sampler that updates the precision matrix $\bm \Omega$ one column at a time.
While the conventional Bayesian methods update $\bm \Omega$ in a one-element-at-a-time manner.
In practice, the value of $\sigma_0^2$ is set to be small, expressing our prior belief that the corresponding $\omega_{ij}$ is 0.
On the other hand, the value of $\sigma_1^2$ is set to be large such that the estimated $\omega_{ij}$ would be very different from 0.
\cite{wang2015scaling} demonstrated that when $\sigma_0 \geq 0.01$ and $\sigma_1 / \sigma_0 \leq 1000$, the MCMC will converge quickly and mix quite well.
Throughout this paper in the numerical study, $\sigma_0$ and $\sigma_1$ are set to be 0.1 and 3, respectively.
Here we consider an informative Beta prior for $\pi_3$ to encourage the sparsity in $\bm \Omega$ by setting $a_5 = q$ and $a_6 = q(q-1)/2$, given the prior belief that $\pi_3$ should be close to 0 to enhance the sparsity of $\bm \Omega$.
We further consider to set hyper-parameter $\lambda=q$ based on the observation from empirical studies that the structures of $\bm \Omega$ are insensitive to a range of $\lambda$.
Note that \cite{wang2015scaling} also suggested to fix $\lambda$ to 5 or 10 if the predictor variables are standardized.


\subsection{Posterior and Inference}
From Formula \eqref{eq:response}, we have the latent vector $\bm \xi_i = \bm{B}^T \bm x_i + \bm{\varepsilon}_i$ with $\bm{\varepsilon}_i \sim \mathcal{N}_q(\bm{B}^{T} \bm{x}_i, \bm \Omega^{-1})$.
Let $\bm{\Xi} = (\bm \xi_1, \bm \xi_2, \ldots, \bm \xi_n)^{T}$ be an $n \times q$ matrix. Based on the priors above, the full-conditional distribution of unknown parameters conditional on the latent variable $\bm \xi$ and data is:
\begin{align*}
 & p(\Tilde{\bm{B}}, \bm{\Omega}, \bm{\tau}, \pi_1, \pi_2, \pi_3, \sigma_{\tau}^2 |\bm{\xi}) \\
    \propto &p(\sigma_{\tau}^2) p(\pi_1) p(\pi_2) p(\pi_3) p(\Tilde{\bm{B}} | \bm{\Omega}, \pi_1) p(\bm{\tau}|\pi_2,\sigma_{\tau}^2) p(\bm{\Omega} | \pi_3)
    \prod_{i=1}^{n}{p(\bm{\xi}_i|\bm{X},\Tilde{\bm{B}},\bm{\Omega},\bm{\tau})}.
\end{align*}
See the full expression in Supplementary Materials A.1.
As a result, the full-conditional distribution of $\Tilde{\bm{B}_g}$ is
\begin{align}\label{post:B}
\vec{(\Tilde{\boldsymbol{B}}^T_g | \mbox{rest} }) \sim
    (1-\pi_{B_g}) \mathcal{N}_{p_gq}(\vec(\bm M_g^T), \bm \Psi_{p_g}\otimes\boldsymbol{\Sigma}) +
    \pi_{B_g} \delta_0(\vec{(\Tilde{\boldsymbol{B}}^T_g)})
\end{align}
for $g=1,\dots,G$, where $\bm \Psi_{p_g} = (\bm I_{p_g} + \bm V_g \mathbb{X}_g^T \mathbb{X}_g \bm V_g)^{-1}$, $\bm M_g = \bm \Psi_{p_g} \bm V_g \mathbb{X}_g^T (\bm{\Xi}-\sum_{k \neq g}^G \mathbb{X}_k \bm V_k \Tilde{\bm{B}}_k)$, and
\begin{align*}
\pi_{B_g} = \frac{\pi_1}{\pi_1 + (1- \pi_1)|\bm \Psi_{p_g}|^{\frac{q}{2}} \exp \left\{ \frac{1}{2}\tr ( \bm{\Omega} \bm{M}_g^T \bm{\Psi}_{p_g}^{-1} \bm{M}_g )
\right\}
}.
\end{align*}
Detailed derivation of $\pi_{B_g}$ is provided in Supplementary Materials A.2.
Denote by $\Tilde{\bm{B}}_{gj}$ the $j$th row of $\Tilde{\bm{B}}_g$,
and $\mathbb{X}_{gj}$ the $j$th column of $\mathbb{X}_g$.
Let $\bm{B}_{-gj}$ represent the matrix $\bm{B}$ without the $j$th row of group $g$, and $\mathbb{X}_{-gj}$ be the corresponding $\mathbb{X}$ without the $j$th column of group $g$.
The full-conditional distribution of $\tau_{g,j}$ is
\begin{align}\label{post:tau}
\tau_{g,j} | \mbox{rest} \sim
    (1-\pi_{\tau_{gj}}) \mathcal{N}^+(\mu_{gj}, \sigma^2_{gj}) +
    \pi_{\tau_{gj}} \delta_0(\tau_{g,j}),
\end{align}
where $\sigma^2_{gj} = [\tr(\bm \Sigma^{-1} \Tilde{\bm{B}}_{gj}^T \mathbb{X}_{gj}^T \mathbb{X}_{gj} \Tilde{\bm{B}}_{gj}) + 1/\sigma_{\tau}^2]^{-1}$, $\mu_{gj} = \sigma^2_{gj} \tr[\bm \Sigma^{-1} (\bm \Xi^T - \bm{B}_{-gj}^T \mathbb{X}_{-gj}^T) \mathbb{X}_{gj} \Tilde{\bm{B}}_{gj}]$, and similarly we have
\begin{align*}
\pi_{\tau_{gj}} &= \frac{p(\tau_{g,j} = 0 | \mbox{rest} )}{p(\tau_{g,j} = 0 | \mbox{rest} ) + p(\tau_{g,j} \neq 0 | \mbox{rest} )}\\
&= \frac{\pi_2}{\pi_2 + 2(1-\pi_2)
(\sigma_{\tau}^2)^{-\frac{1}{2}} (\sigma_{gj}^2)^{\frac{1}{2}} \exp \left\{
\frac{1}{2} \frac{\mu_{gj}^2}{\sigma_{gj}^2}\right\}
 \Phi \left( \frac{\mu_{gj}}{\sigma_{gj}}\right)
},
\end{align*}
where $\Phi(\cdot)$ is the cumulative distribution function for the standard normal variable.

The full-conditional distributions for $\pi_1$, $\pi_2$, $\pi_3$ and $\sigma_{\tau}^2$ are
\begin{align}
\pi_1 | \mbox{rest} &\sim \mbox{Beta} \left( a_1 + \sum_{g=1}^G \bm I(\Tilde{\bm{B}}_g = \bm{0}), a_2 + \sum_{g=1}^G \bm I(\Tilde{\bm{B}}_g \neq \bm{0}) \right),  \label{post:pi1} \\
\pi_2 | \mbox{rest} &\sim \mbox{Beta} \left( a_3 + \sum_{g=1}^G \sum_{j=1}^{p_g} \bm I(\tau_{g,j} = 0), a_4 + \sum_{g=1}^G \sum_{j=1}^{p_g} \bm I(\tau_{g,j} \neq 0) \right),  \label{post:pi2} \\
\pi_3 | \mbox{rest} &\sim \mbox{Beta} \left( a_5 + \sum_{i<j} \bm I(\omega_{ij} \neq 0), a_6 + \sum_{i<j} \bm I(\omega_{ij} = 0) \right),  \label{post:pi3} \\
\sigma_{\tau}^2 | \mbox{rest} &\sim \mbox{IG} \left(1 + \frac{1}{2} \sum_{g=1}^G \sum_{j=1}^{p_g} \bm I(\tau_{g,j} \neq 0), d + \frac{1}{2} \sum_{g=1}^G \sum_{j=1}^{p_g} \tau_{g,j}^2 \right). \label{post:ssquared}
\end{align}
Next we examine the posterior distribution of $\bm \Omega$.
To facilitate the expression, we introduce latent variables $z_{ij}$ and re-write the prior \eqref{prior:Omega} as
\begin{align*}
p(\bm \Omega) | & \propto
    \prod_{i<j} \mathcal{N} (\omega_{ij} ; 0, \sigma_{z_{ij}}^2) \prod_{i}^p \mbox{e} (\omega_{ii} ; \frac{\lambda}{2}), \\
    p(z_{ij}) &= \pi_3^{z_{ij}} (1 - \pi_3)^{1-z_{ij}},
\end{align*}
where $z_{ij} = 0$ or 1 according to whether $\omega_{ij} = 0$ or not.
Hence the variable $z_{ij}$ follows Bernoulli distribution with parameter $\pi_3$.
Now the full-conditional distribution of $\bm \Omega$ is
\begin{align}\label{post:Sigma}
&p(\bm \Omega | \mbox{rest}) \propto |\bm \Omega|^{\alpha/2} \exp\{ -\frac{1}{2} \tr (\bm \Omega \bm \Theta) \} \prod_{i<j} \left \{ \exp(-\frac{\omega_{ij}^2}{2 \sigma_{z_{ij}}^2}) \right \} \prod_{i}^{p} \exp \left(-\frac{\lambda}{2} \omega_{i i}\right), \\
&p(z_{ij}=1 | \mbox{rest}) = \frac{\mathcal{N}\left(\omega_{i j} ; 0, v_{1}^{2}\right) \pi_3}{\mathcal{N}\left(\omega_{i j} ; 0, v_{1}^{2}\right) \pi_3+\mathcal{N}\left(\omega_{i j} ; 0, v_{0}^{2}\right)(1-\pi_3)},  \nonumber
\end{align}
where $\alpha = n + \sum_{g=1}^G p_g \bm I(\Tilde{\bm{B}}_g \neq \bm{0})$ and $\bm \Theta = (\bm{\Xi}-\mathbb{X} \bm{B})^T (\bm{\Xi}-\mathbb{X} \bm{B}) + \Tilde{\bm{B}}^T \Tilde{\bm{B}}$.
Sampling $\bm \Omega$ from its posterior \eqref{post:Sigma} adopts the procedures described here:

Let $\bm H = (\sigma_{z_{ij}}^2)_{p \times p}$ be a symmetric matrix with zeros as its diagonal entries and $(\sigma_{z_{ij}}^2)_{i < j}$ as its upper diagonal entries.
Partition $\bm \Omega$, $\bm \Theta$ and $\bm H$ as
\begin{align*}
\bm \Omega = \left[
\begin{array}{cc}
\bm \Omega_{11},      & \bm \varphi_{12} \\
\bm \varphi_{12}',   & \varphi_{22}
\end{array}\right],~~~
\bm \Theta = \left[
\begin{array}{cc}
\bm \Theta_{11},      & \bm \theta_{12} \\
\bm \theta_{12}',   & \theta_{22}
\end{array}\right],~~~ \mbox{and} ~~~
\bm H = \left[
\begin{array}{cc}
\bm H_{11},      & \bm h_{12} \\
\bm h_{12}',   & 0
\end{array}\right].
\end{align*}
Consider a variable transform: $(\bm \varphi_{12}, \varphi_{22}) \rightarrow (\bm \eta = \bm \varphi_{12}, \zeta = \varphi_{22} - \bm \varphi_{12}' \bm \Omega_{11}^{-1} \bm \varphi_{12})$.
Then the conditional distributions of $\bm \eta$ and $\zeta$ are
\begin{align*}
\bm \eta | \mbox{rest} \sim \mathcal{N} (-\bm \Sigma_{\eta} \bm \theta_{12}, \bm \Sigma_{\eta}), ~~~ \mbox{and}~~~ \zeta | \mbox{rest} \sim \mbox{Gamma} (\frac{\alpha}{2} + 1, \frac{\theta_{22}+\lambda}{2}),
\end{align*}
where $\bm \Sigma_{\eta} = ( (\theta_{22} + \lambda) \bm \Omega_{11}^{-1} + \mbox{diag}(\bm h_{12})^{-1})^{-1}$.

To construct matrix $\bm{\Xi} = (\bm \xi_1, \bm \xi_2, \ldots, \bm \xi_n)^{T}$ in the posteriors, we sample $\bm \xi_i$ according to
\begin{align}\label{eq:xi}
f(\bm \xi_i \mid \bm y_i, \bm B, \bm \Sigma)
\propto& |\bm \Omega|^{\frac{1}{2}} \exp \left( -\frac{1}{2}[\bm{\xi}_i - \bm{B}^{T} \bm{x}_i]^T \bm \Omega [\bm{\xi}_i - \bm{B}^{T} \bm{x}_i] \right) \prod_{j=1}^{l} \frac{1}{\sqrt{2 \pi} \sigma^{(j)}} \exp{\{-\frac{(u_i^{(j)} - \mu^{(j)})^2}{2 {\sigma^{(j)}}^{2}}\}}  \nonumber \\
& \prod_{j=1}^{m} \frac{(\lambda^{(j)})^{z_i^{(j)}} \exp{(-\lambda^{(j)})}}{z_i^{(j)} !} \cdot \prod_{j=1}^{k} (\gamma^{(j)})^{w_i^{(j)}} (1 - \gamma^{(j)})^{1 - w_i^{(j)}},
\end{align}
where a non-informative prior $\mbox{IG}(1/2,1/2)$ is assumed for the parameter ${\sigma^{(j)}}^2$, such that it can be sampled from the conditional distribution
\begin{align}\label{eq:sigmaj}
{\sigma^{(j)}}^2 | \mbox{rest} \sim \mbox{IG}\left(\frac{1}{2}+n,\frac{1}{2}+\frac{1}{2}\sum_{i=1}^n (u_i^{(j)}-\bm \xi_i^{(j)})\right),
\end{align}
where $\bm \xi_i^{(j)}$ is the $j$th element of $\bm \xi_i$.
See a derivation of Equation~\eqref{eq:xi} in Supplementary Materials A.3.
Therefore, the Gibbs sampling for the proposed BS-MRMR model is summarized in Algorithm \ref{alg:gibbs}.
\begin{algorithm}[H]
\caption{Gibbs sampling for BS-MRMR model}
\label{alg:gibbs}
\begin{algorithmic}

\Repeat
\State {Sample $\bm{\xi}_i$ by $f(\bm \xi_i \mid \bm y_i, \bm B, \bm \Sigma)$ from \eqref{eq:xi} and \eqref{eq:sigmaj}.}

\State {Sample $\Tilde{\bm B}_g$ from \eqref{post:B}.}

\State {Sample $\tau_{g,j}$ from \eqref{post:tau}, and compute $\bm B_g$}.

\State {Sample $\pi_1$, $\pi_2$, $\pi_3$ and $\sigma_{\tau}^2$ from \eqref{post:pi1} to \eqref{post:ssquared}.}

\State {Sample $\bm \Omega$ from \eqref{post:Sigma}.}

\Until {Convergence}

\end{algorithmic}
\end{algorithm}

\section{Numerical Study} \label{sec:simu}
In this section, we evaluate the performance of the proposed model, denoted as BS-MRMR, by comparison with separate models 1) BS-GLM, 2) FS-GLM, and 3) a hierarchical generalized transformation (HGT) model (\citealt{bradley2022joint}).
The BS-GLM method separately fits each response using the Bayesian generalized linear model, with its variable selection conducted according to the 95\% credible intervals.
The implementation of the BS-GLM method is conducted by $bayesglm(\cdot)$ function in the R software.
The FS-GLM method separately fits each response on all the predictor variables via generalized linear model using the Lasso regularization.
Precisely, the continuous, counting and binary responses are fitted through linear, Poisson and logistic regressions with Lasso penalties, respectively.
This is implemented by $glmnet(\cdot)$ function in the R software.
The third benchmark HGT model was recently developed to transform mixed responses into continuous responses (\citealt{bradley2022joint}).
The HGT is a two-step model, where the first step samples latent continuous variables for the mixed responses, and the second step estimates a Bayesian multivariate regression model for each sample.
The original article suggested to use a Bayesian mixed effects model as the second step estimator.
However, such a model does not consider group and individual sparsity.
For fair comparison,
a sparse multivariate regression model called MBSGSSS (\citealt{liquet2017bayesian}) was selected to be the second step model here.
In the rest of this manuscript, we denote the HGT-MBSGSSS model as HGT for short.

Regarding the dependency among multiple responses, we consider the following matrix structures of $\bm \Omega = (\omega_{ij})_{q \times q}$.
\begin{itemize}
\item Scenario 1 $\bm \Omega_1: \omega_{ij}=1_{\{i=j\}} + 0.5_{|i-j|=1}+0.3_{|i-j|=2}+0.1_{|i-j|=3}$.
\item Scenario 2 $\bm \Omega_2$ is generated by randomly permuting rows and corresponding columns of $\bm \Omega_1$.
\item Scenario 3 $\bm \Omega_3:\omega_{ij}=0.5^{|i-j|}$.
\item Scenario 4 $\bm \Omega_4$: randomly and evenly divide the indices $1,2, \ldots, q$ into $M$ groups. Let $\omega_{jj} = 1$. Set $\omega_{jk} = 0.4$ for $j \neq k$ if $j$ and $k$ belong to the same group and 0 otherwise.
\item Scenario 5 $\bm \Omega_5 = \left(
\begin{array}{cc}
    \mbox{CS}(0.5) & \bm 0 \\
    \bm 0 &  \bm I
  \end{array}
\right)$, where CS(0.5) represents a $4 \times 4$ compound structure matrix with diagonal elements 1 and others 0.5.
$\bm 0$ indicates a matrix with all elements 0.
\end{itemize}

Scenario 1 is a banded matrix representing each response is only correlated with its several nearest responses.
Scenario 2 disrupts such sparse structure but maintaining the same sparse extent.
Scenario 3 is an autoregressive model with its elements decaying as one moves away
from the diagonal.
Scenario 4 is a random sparse matrix.
Scenario 5 represents that the first 4 responses are correlated but independent from others.
We generate $n=100$ training data to estimate the model and 100 testing data to examine the model performance, both of which are from the multivariate normal distribution $\mathcal{N}_p(0, \sigma_X \bm I)$ with (1) $p=20, l=m=k=2$ and (2) $p=80, l=m=k=5$.
Here we choose $n=100$ as the training sample size to stress the proposed model with a large number of parameters in comparison with the sample size.
For example, $p=20, l=m=k=2$ results in $(p+1)(l+m+k)=126$ linear model coefficients to be estimated including the intercepts; and $p=80, l=m=k=5$ results in $(p+1)(l+m+k)=1215$ linear model coefficient parameters.
We also we consider a simulation setting with $n=50, p=80$ to further evaluate the proposed model. The results are summarized in the Supplemental Materials (Tables 5-7).

When $p=20$, we take $M=3$ for Scenario 4,
and the coefficient matrix is divided into 4 groups with each group having 5 variables as $\bm B_1^T = (\ast \ast~0 0 \ast~\bm 0_5~\ast \ast~0 0 \ast~\bm 0_5)$ and $\bm B_2^T = (\bm 0_5~\ast \ast~0 0 \ast~\ast \ast~0 0 \ast~\bm 0_5)$, where $\ast$ represents that the corresponding columns are not zeros generating from uniform distribution Unif$(l_B, u_B)$, $0$ represents that the corresponding columns are zeros, and $\bm 0_a$ represents that the corresponding $a$ columns are zeros.
When $p=80$, the value of $M$ is set to be 5 in Scenario 4.
We consider 6 groups of predictor variables and divide coefficient matrix $\bm B_3^T = ((\bm 0_5, \bm \ast_5) ~ \bm 0_{20} ~(\bm \ast_5, \bm 0_5) ~ \bm 0_{10} ~ (\bm \ast_5, \bm 0_{15}) ~ \bm 0_{10})$ as well as $\bm B_4^T = (\bm 0_{20} ~ (\bm 0_5, \bm \ast_5) ~ (\bm \ast_5, \bm 0_5) ~ (\bm \ast_5, \bm 0_{15}) ~ \bm 0_{10} ~ \bm 0_{10})$.
The observations of the response variables are then generated based on Formula \eqref{eq:response}.
The data generation parameters $\sigma_X$, $l_B$ and $u_B$ are tuned
to make sure that the counting observations in the response matrix for each setting are within a reasonable range.
We generate 10000 MCMC samples with the first 2000 draws as burn-in period.
The median values of the rest 8000 samples are taken as the parameter estimates for the proposed model.

\begin{figure}[H]
\begin{center}
\scalebox{0.6}[0.6]{\includegraphics{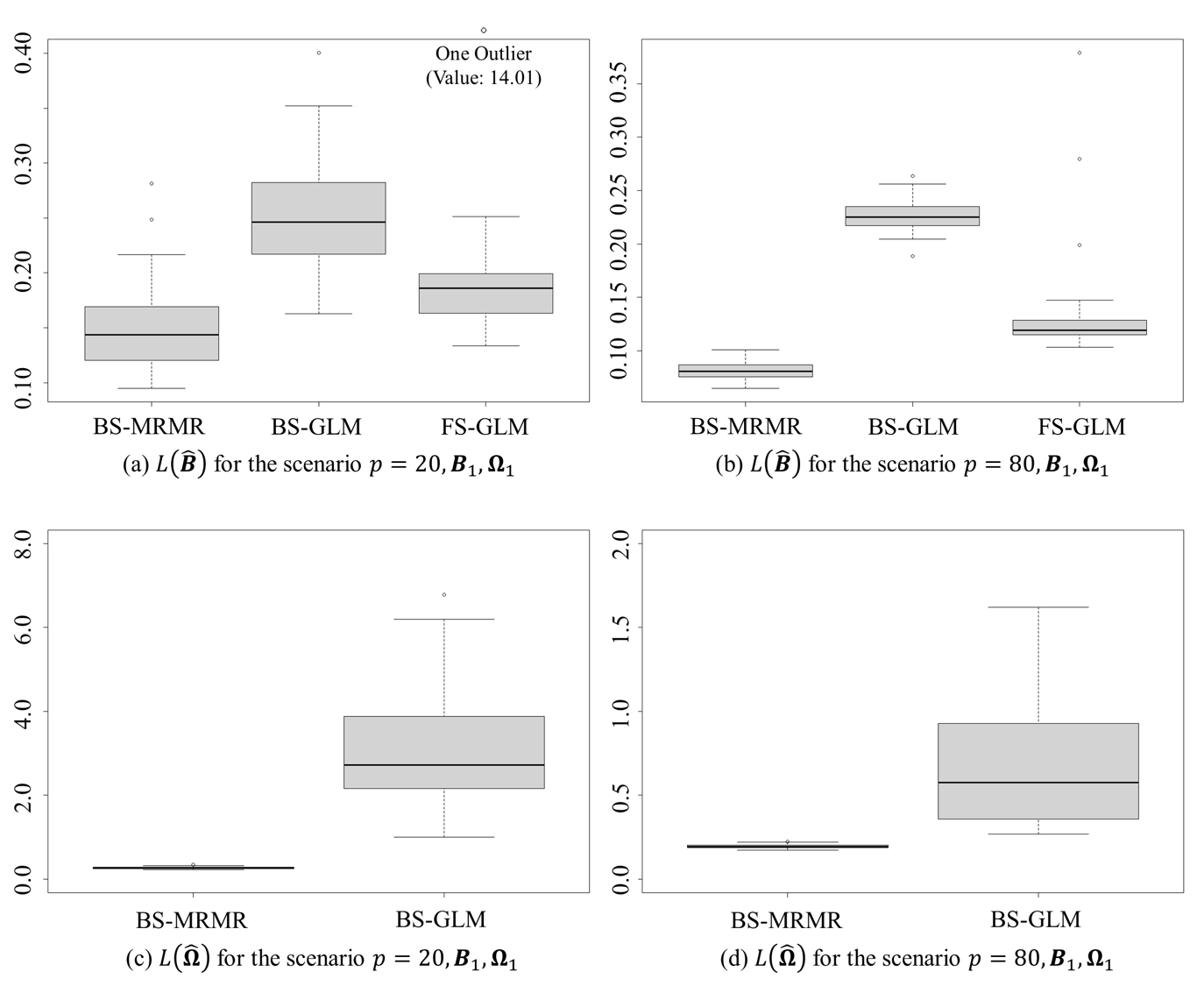}}
\caption{Box plots of two selected scenarios for parameter estimation errors of $\hat{\boldsymbol{B}}$ and $\hat{\boldsymbol{\Omega}}$ to compare BS-MRMR (proposed), FS-GLM, and BS-GLM. See full results in supplementary materials (Table 1). }\label{Fig:tbl1}
\end{center}
\end{figure}

\begin{figure}[H]
\begin{center}
\scalebox{0.6}[0.6]{\includegraphics{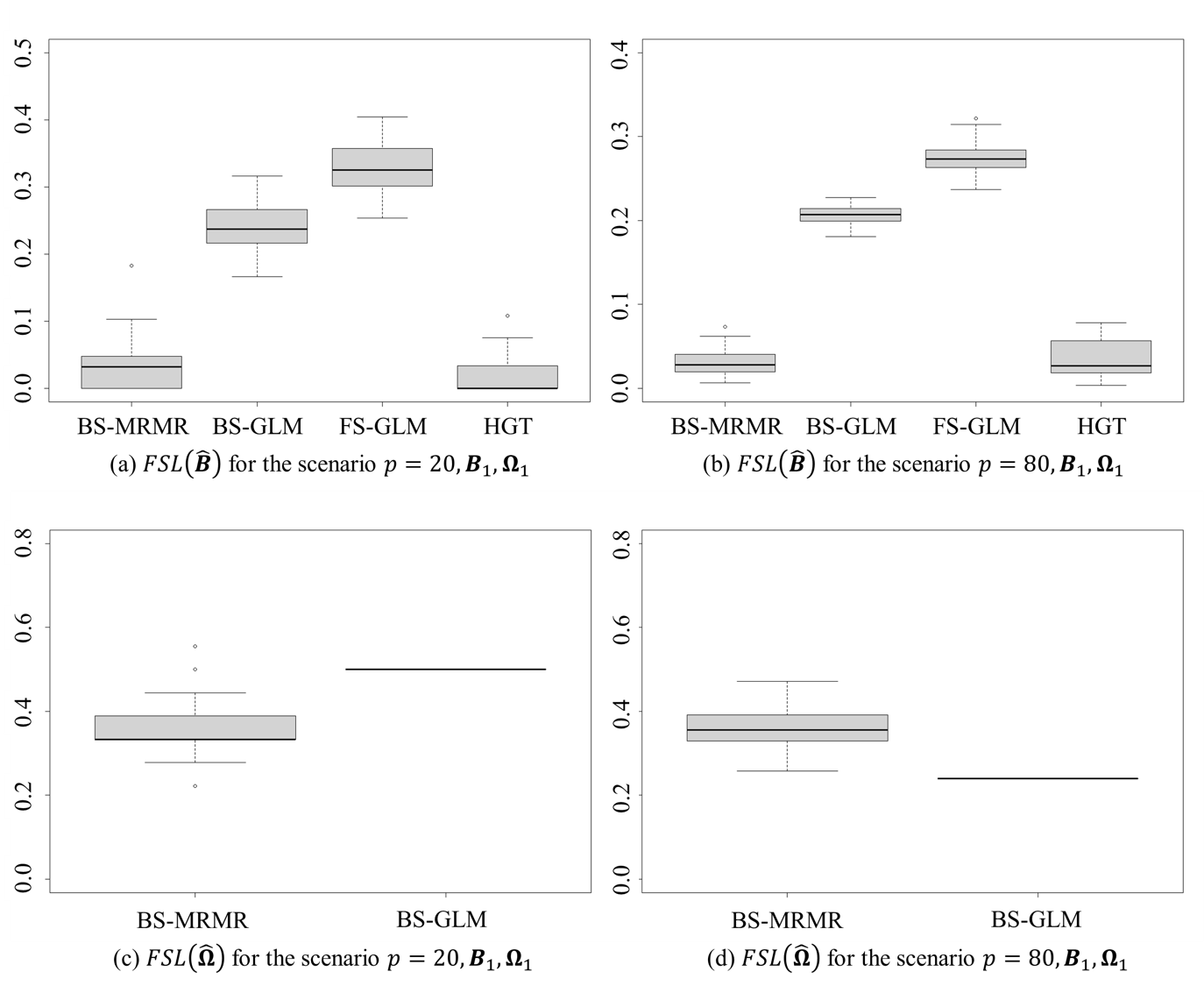}}
\caption{Box plots of two selected scenarios for variable selection errors of $\hat{\boldsymbol{B}}$ and $\hat{\boldsymbol{\Omega}}$ to compare BS-MRMR (proposed), FS-GLM, BS-GLM, and HGT. See full results in supplementary materials (Table 2). }\label{Fig:tbl2}
\end{center}
\end{figure}

To evaluate the accuracy of model estimation, we consider the loss measures
\begin{align*}
L(\hat{\bm B}) = \sqrt{\frac{\sum_{i=1}^p \sum_{j=1}^q (\boldsymbol{B}_{ij} - \hat{\boldsymbol{B}}_{ij})^2}{pq}}  ~~~ \mbox{and} ~~~ L(\hat{\bm \Omega}) = \sqrt{\frac{\sum_{i=1}^q \sum_{j=1}^q (\boldsymbol{\Omega}_{ij} - \hat{\boldsymbol{\Omega}}_{ij})^2}{q^2}},
\end{align*}
where $\hat{\bm B}$ and $\hat{\bm \Omega}$ denote the estimates of matrices $\bm B$ and $\bm \Omega$.
For gauging the performance of variable selection in $\hat{\boldsymbol{B}}$ and sparsity imposed in $\hat{\boldsymbol{\Omega}}$, we also consider $FSL = $ false positive (FP) + false negative (FN).
Figures~\ref{Fig:tbl1} and \ref{Fig:tbl2} summarize via boxplots the results of two selected scenarios of these loss measures for each method over 50 replicates.
Note that parameter estimation errors of the HGT method are not reported in Figure~\ref{Fig:tbl1} since the HGT method transforms responses into different scales, making it non-comparable with other methods.
Please refer to Tables 1 and 2 in Supplementary Materials for the full comparison results.

It is clear to see that the proposed BS-MRMR model generally outperforms other compared methods for all the settings with respect to loss measures $L(\hat{\bm B})$, $L(\hat{\bm \Omega})$, $FSL(\hat{\bm B})$, and $FSL(\hat{\bm \Omega})$.
The proposed model produces the lowest values of losses and corresponding standard errors, because it takes advantage of the association between responses and model them jointly.
Similar conclusion can be readily drawn by comparing the HGT model with two separate GLM benchmark models.
The main reason is that the BS-GLM and FS-GLM methods fit data separately, losing the potential information of the response variables' relationship.
Specifically, when the number of predictor variables $p=20$, the BS-GLM method sometimes performs better than the FS-GLM for the loss $L(\hat{\bm B})$, but other times it is worse, depending on the structures of $\bm \Omega$ and coefficient matrices $\bm B$.
When $p$ increases to 80, the FS-GLM method appears to be better regarding the loss $L(\hat{\bm B})$. Nevertheless, they are all inferior to the proposed model.
For the loss $L(\hat{\bm \Omega})$, note that the separate modeling FS-GLM cannot provide the correlation among responses.
And the metric $FSL(\hat{\bm \Omega})$ for HGT is not accessible since the function ``MBSGSSS'' provided by \cite{liquet2017bayesian} does not return samples for $\hat{\bm \Omega}$.
In addition, regarding the loss $FSL(\hat{\boldsymbol{B}})$, the proposed BS-MRMR model is substantially superior over the compared methods, implying that it is able to accurately identify the group sparsity and individual sparsity within each group.
We also observe that the BS-GLM method performs better than the FS-GLM with respect to $FSL(\hat{\boldsymbol{B}})$.
It is also interesting to observe that the HGT model can have a comparable performance with the proposed BS-MRMR model in terms of $FSL(\hat{\boldsymbol{B}})$.
Note that, in the current implementation of the HGT method, we adopted the default settings of hyperparameters in the supplemented programming code in \cite{bradley2022joint}.
When comparing the loss $FSL(\hat{\boldsymbol{\Omega}})$, the proposed BS-MRMR model does not consistently outperform the BS-GLM, especially when the number of responses increases from 6 to 15.
The reason is that the BS-GLM only provides estimates for diagonal elements in $\hat{\boldsymbol{\Omega}}$, hence the loss $FSL(\hat{\boldsymbol{\Omega}})$ tends to be smaller as the underlying ${\boldsymbol{\Omega}}$ becomes sparser.

\begin{figure}[H]
\begin{center}
\scalebox{1}[1]{\includegraphics{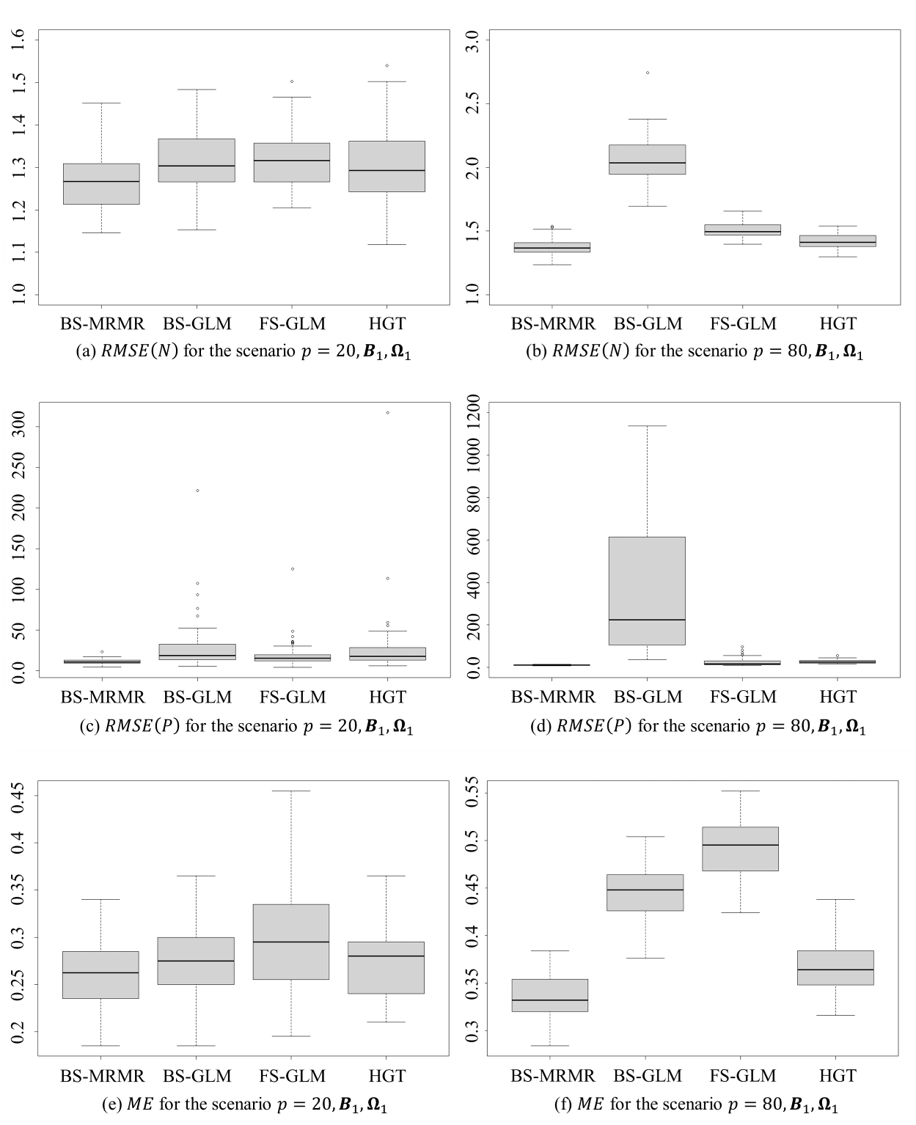}}
\caption{Box plots of two selected scenarios for prediction errors to compare BS-MRMR (proposed), FS-GLM, BS-GLM, and HGT. See full results in supplementary materials (Tables 3-4). }\label{Fig:tbl34}
\end{center}
\end{figure}

To further investigate the prediction performance of the proposed model, the root-mean-square error (RMSE) $\sqrt{\sum_{i=1}^{100}(y_i - \hat{y}_i)^2 / 100}$ is computed for the continuous and counting responses via the testing data, denoted by $RMSE(N)$ and $RMSE(P)$, where $\hat{y}_i$ stands for the corresponding fitted value.
Naturally, we use the misclassification error rate ($ME$) $\sum_{i=1}^{100}\bm I(y_i \neq \hat{y}_i) / 100$ to compare the model performance on the binary response.
The cut-off point for the binary response estimates is 0.5.
Figure~\ref{Fig:tbl34} reports the prediction results of the selected scenarios for the methods in comparison.
Please see the full comparison results in the Tables 3 and 4 in Supplementary Materials.
We observe that the proposed BS-MRMR model provides more accurate predictions on continuous, counting and binary responses for all the settings.
Specifically, it is slightly better than the compared methods with respect to $RMSE(N)$, and substantially better in terms of $RMSE(P)$ and $ME$.
When the number of predictor variables increases from 20 to 80, the proposed model performs consistently well with the lowest loss values and standard errors.
Such results demonstrate that incorporating the association of responses in the proposed model will obviously improve its prediction performance.
In addition, we also observe that the FS-GLM and BS-GLM methods occasionally have convergence issue when fitting the counting responses, yielding high values of $RMSE(P)$ especially for larger $p=80$.
This is possibly due to some very large values of counting responses in the datasets, which remarkably increases the difficulty of modeling and hence causes convergence issue.
In contrast, the relatively lower values of $RMSE(P)$ produced by the proposed method demonstrate that the BS-MRMR model is more robust than the compared approaches for the underlying multivariate datasets.

\begin{figure}[h]
\begin{center}
\scalebox{0.5}[0.5]{\includegraphics{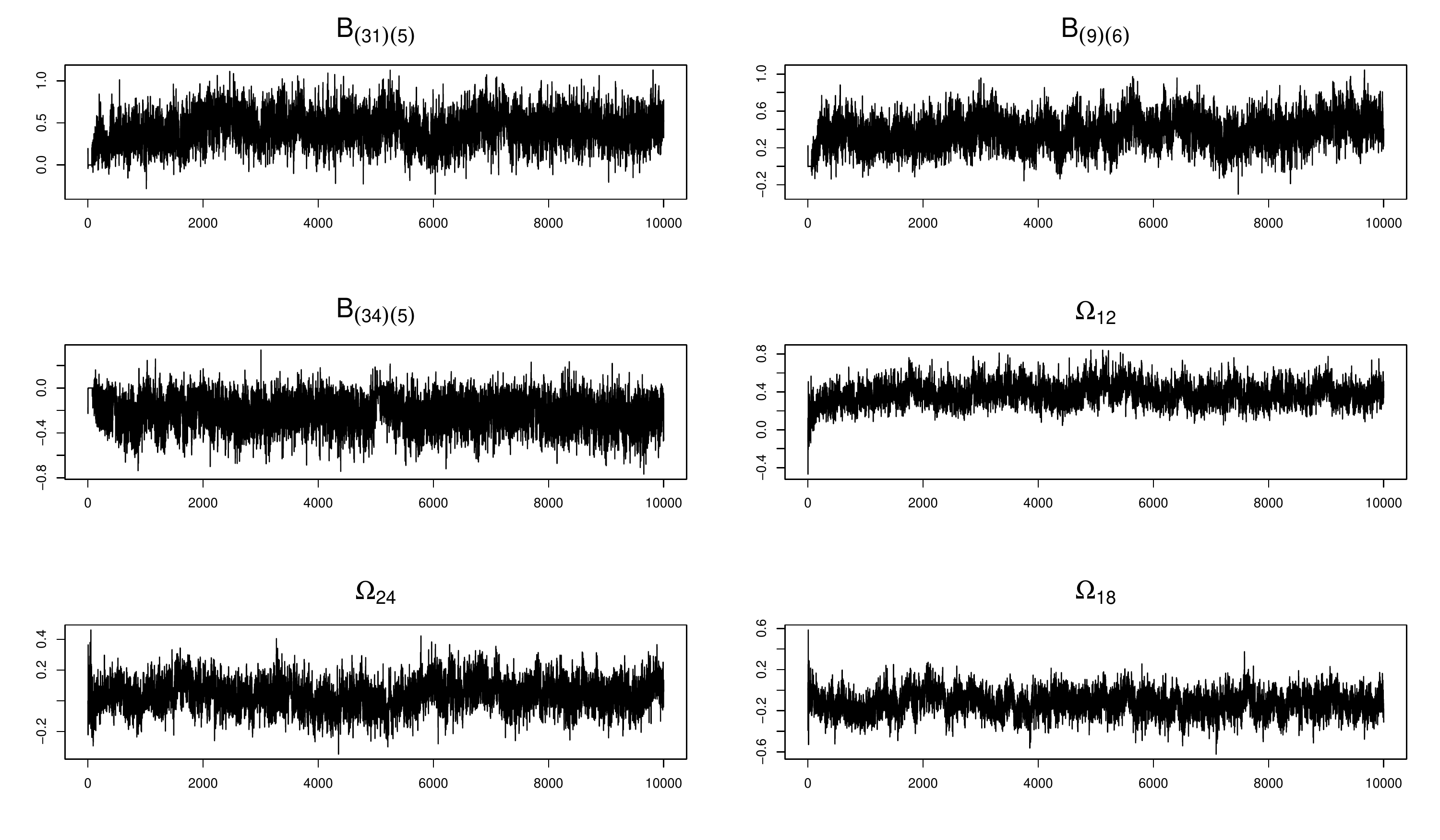}}
\caption{Trace plots for selected parameters from one replicate for Scenario 1 and $\bm B_3$ when $p=80$.}\label{Fig:trace.sim}
\end{center}
\end{figure}

\begin{figure}[h]
\begin{center}
\scalebox{0.5}[0.5]{\includegraphics{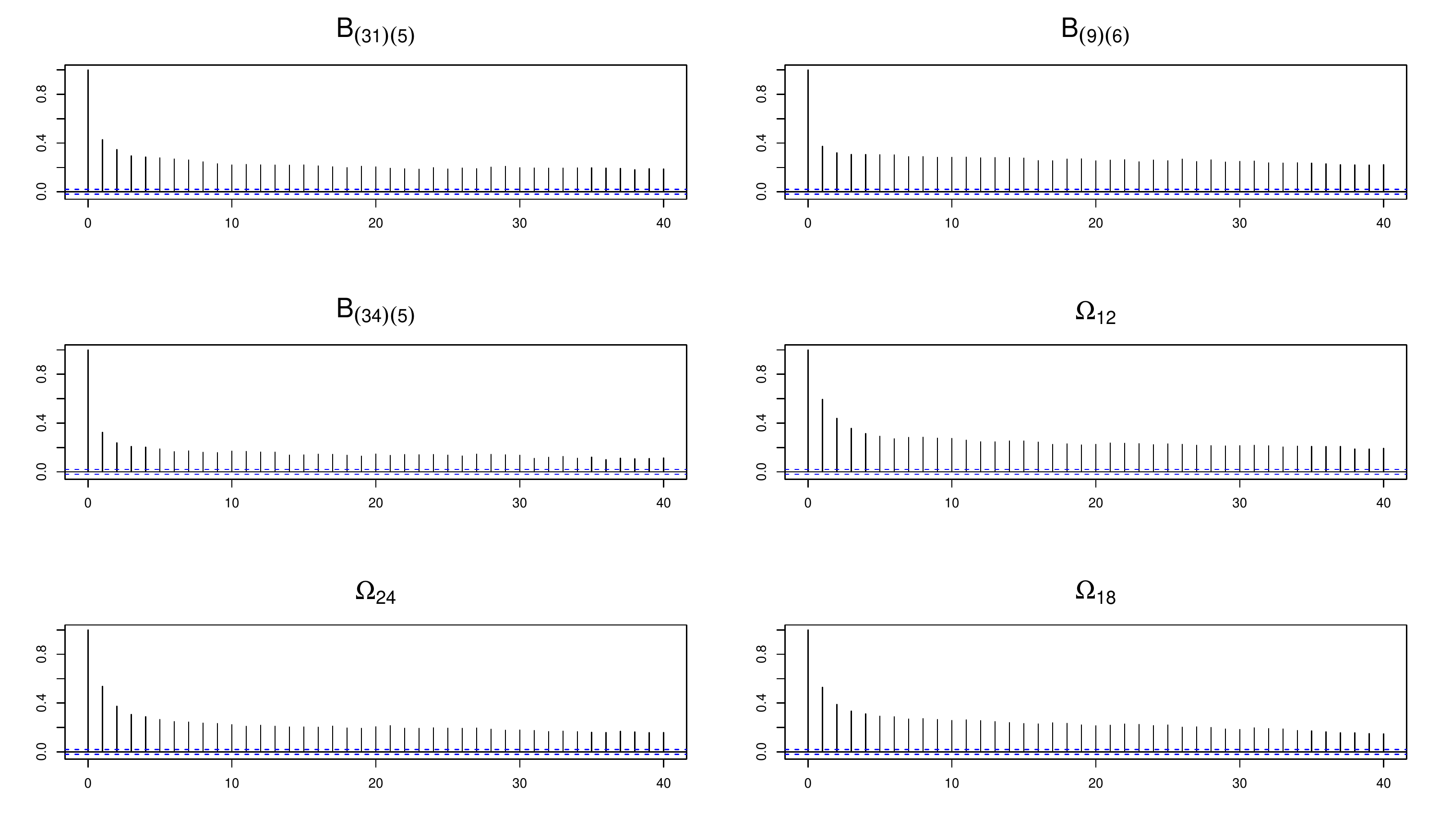}}
\caption{ACF plots for selected parameters from one replicate for Scenario 1 and $\bm B_3$ when $p=80$.}\label{Fig:acf.sim}
\end{center}
\end{figure}

Figure \ref{Fig:trace.sim} shows the trace plots of randomly selected parameters in the precision matrix $\bm \Omega$ and coefficient matrix $\bm B$ from one replicate for Scenario 1 and $\bm B_3$ when $p=80$.
It is seen that the traces of the parameters fluctuate around the means with relatively stable variation, indicating that the MCMC chains converge.
Figure \ref{Fig:acf.sim} displays the corresponding autocorrelation functions of those parameters.
The quick decrease of ACF in these plots implies the fast convergence of the Gibbs sampling iterations.
The rest of the parameters present the similar patterns, and hence their plots are omitted.
We further compare the computation time for each method for the simulation study ($p=20, \boldsymbol{B}_1, \boldsymbol{\Omega}_1$).
The average time for model estimation and prediction are summarized in Table~\ref{table:computation_time}.
We note that the long computation time of HGT could be due to the current implementation of the two-step estimation procedure.
Specifically, the implementation of HGT method firstly samples latent continuous variables for the mixed responses, then estimates a Bayesian multivariate regression model for each sample.
Consequently, the estimation time of HGT is linearly related to the number of samples obtained from the first step.
It is worth to remarking that there could be a computationally more efficient implementation by combining the Gibbs sampler of HGT with the second step sampler \citep{bradley2022joint}.

    \begin{table}[H]
    \begin{center}
    \caption{{The averages and standard errors (in parenthesis) of computation time for the numerical study with $p=20, \boldsymbol{B}_1, \boldsymbol{\Omega}_1$.}}
    \label{table:computation_time}
    \begin{tabular}{cccccccc}
    \hline \hline
    &    & Estimation Time (s)  & Prediction Time (s)   \\
    \hline
    &FS-GLM     &334.9 (5.854) &0.158 (0.004)       \\
    &BS-GLM     &0.035 (0.000) &0.001 (0.000)        \\
    &HGT        &67652 (108.5) &4.414 (0.041) \\
    &BS-MRMR    &0.059 (0.001) &0.002 (0.000)       \\
    \hline \hline
    \end{tabular}
    \end{center}
    \end{table}

\section{Case Study in Fog Manufacturing} \label{sec:case}
\begin{figure}
\begin{center}
\scalebox{0.55}[0.55]{\includegraphics{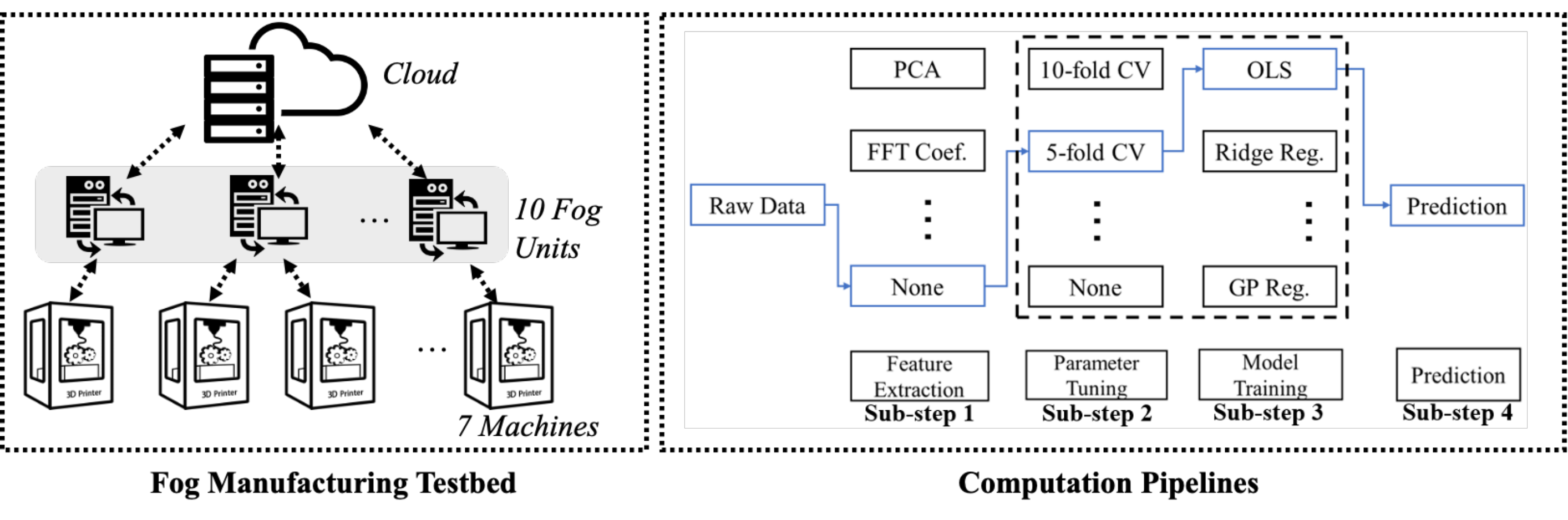}}
\caption{Fog manufacturing testbed and recommended computation pipelines. Redrawn from \cite{zhang2019fog} with authors' permission. }\label{Fig:fog_mfg}
\end{center}
\end{figure}

This section investigates a real case study in a Fog manufacturing testbed for evaluating the proposed BS-MRMR model.
The three-layer architecture of this testbed is presented in the left panel of Figure~\ref{Fig:fog_mfg}.
In the top layer, a central computation unit (CPU: i7-6700k) serves as the orchestrator to master the offloading of computation tasks and collect the results from each Fog unit.
In the middle layer, ten Raspberry Pi 3 devices are deployed as Fog nodes with different computation capabilities and communication bandwidths.
In the bottom layer, seven manufacturing machines are connected to each Fog nodes, such that the manufacturing process data from any machine can be collected by each of ten Fog nodes.
All the runtime performance signals (i.e., CPU and memory utilizations, temperature of the Fog nodes, etc.) during the execution are stored locally in each Fog nodes based on a Python program.
For the computation tasks to be offloaded, the right panel of Figure~\ref{Fig:fog_mfg} presents the computation pipelines with four sub-steps and multiple method options in each sub-step following the definition described in \cite{chen2020adapipe}.
Here, each option from one sub-step is treated as a computation task, and one sequence from sub-step 1 to sub-step 4 represents a computation pipeline.
The predictive offloading methods aim at dynamically assigning these computation tasks into different Fog nodes considering the responsiveness and reliability as detailed in \cite{chen2018predictive}.
The offloading decisions are made based on the prediction of the runtime performance metrics extracted from the runtime performance signals.

To generate the runtime performance signals for analysis,
an experiment of four factors with two levels was conducted by the full factorial design in Table~\ref{table:doe}.
There are 32 runs in total executed with two replicates for each treatment.
The workflow of this Fog manufacturing testbed in this experiment follows three steps:
first, the computation pipelines to be offloaded in sub-steps are selected following two task selection strategies, namely, (1) random selection from all candidate pipelines, and (2) recommendation-based selection to choose the Top-ranked pipelines suggested by a recommender system (see \cite{chen2020adapipe} for details);
Second, the orchestrator then provides the offloading decisions (randomly or following a time-balanced offloading strategy) to assign sub-steps of the selected computation pipelines to different Fog nodes for execution;
Finally, Fog nodes check whether the dataset to support the assigned sub-steps exists in their local storage according to the data storage strategy.
A Fog unit will download the necessary dataset from other Fog nodes or the orchestrator, then will execute the assigned sub-step with runtime performance metrics recorded.

\begin{table}[H]
\begin{center}
\caption{Design of experiments for the analysis of runtime performance metrics (summarized from \cite{zhang2019fog} with authors' permission).}
\label{table:doe}
\resizebox{\textwidth}{!}{ 
\begin{tabular}{cc|cc}
\hline \hline
& {\bf Design Factors}    & {\bf Level 1}  & {\bf Level 2}   \\
\hline
&Task selection strategy &Random selection &Recommendation (\citealt{chen2020adapipe})  \\
&Number of pipelines &5 &10  \\
&Data storage strategy  &One copy on each Fog node &3 copies randomly stored on 3 Fog nodes \\
&Offloading strategy  &Random offloading &Time-balanced offloading \\
\hline \hline
\end{tabular}}
\end{center}
\end{table}

In this case study, the observational data are $\{\boldsymbol{x}_{t_{f}}, \boldsymbol{Y}_{t_{f}}\}$, where $t_f=1,\dots,n_f$ is defined as the $t_f$-th sub-step that is assigned to the $f$-th Fog unit, $f=1,\dots,10$.
The $\boldsymbol{x}_{t_{f}^{(1)}}\in \mathbb{R}^{48}$ is the predictor vector that contains two groups of features (i.e., Group 1: 11 summary statistics of each of the three runtime performance signals; and Group 2: 17 dummy variables as the embedding of the assigned sub-step) from a previous time instance.
The $\boldsymbol{Y}_{t_{f}^{(2)}}\in \mathbb{R}^{5}$ is the response vector which contains five runtime performance metrics in mixed types when executing the $t_f$-th sub-step, i.e., continuous metrics: averaged CPU utilization $\mathbb{Y}_1$, averaged temperature $\mathbb{Y}_2$; counting metric: number of sub-steps that can be executed within five seconds $\mathbb{Y}_3$; and binary metrics: whether temperature exceeds a certain threshold $\mathbb{Y}_4$, and whether memory utilization exceeds a certain threshold $\mathbb{Y}_5$.
In total, 3407 samples were obtained from the total 10 Fog nodes. Each Fog unit has around 340 samples ordered by timestamps when the samples are collected.
The training data consists of the first 200 samples from each Fog unit, and the testing data consists of the remaining samples from each Fog unit.

We compare the proposed BS-MRMR model with the FS-GLM, BS-GLM, and HGT model.
For the BS-MRMR, the values of hyperparameters of priors are the same as those in Section 4 to encourage both group and individual sparsity for predictor variables.
In addition, 10000 MCMC samples from the proposed model are drawn with the first 2000 as burn-in period.
For the HGT method, we adopted the default settings of hyperparameters in the supplemented programming code in \cite{bradley2022joint}.
For this real data of Fog manufacturing, there can be $12$ hyperparameters in the HGT method.
It would be very challenging to carefully adjust hyperparameters to avoid the bias issue introduced by transformation.
One possibility may specify hyperparameters for the HGT methods using certain Bayesian optimization techniques when the number of hyperparameters is large.

\begin{figure}[H]
\begin{center}
\scalebox{0.82}[0.82]{\includegraphics{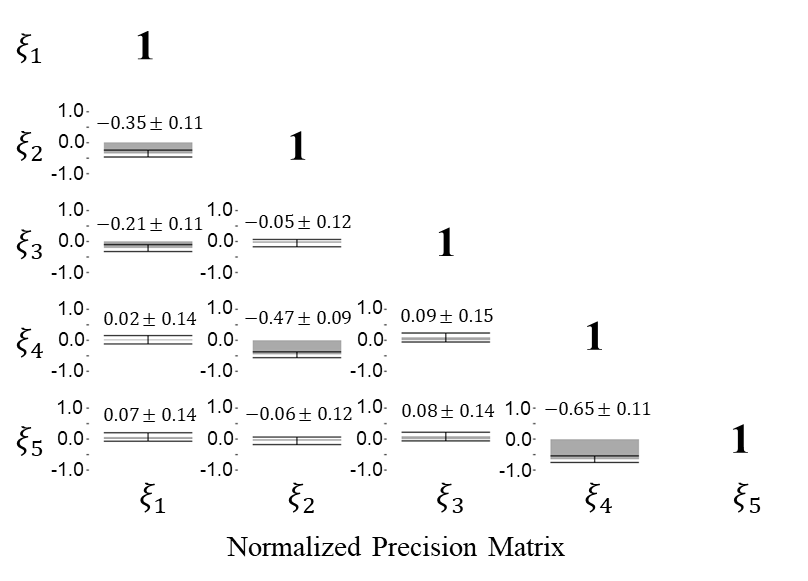}}
\caption{Estimated dependency (i.e., precision matrix) of the latent responses (i.e., $\xi_1, \cdots, \xi_5$) from the BS-MRMR method, where the shadow bars present the median values, and 95\% credible intervals are presented in the form of error bars and labels on subplots. }\label{Fig:omega_interval}
\end{center}
\end{figure}

The averages and standard errors of loss measures $RMSE(N)$, $RMSE(P)$ and $ME$ over all ten Fog nodes are summarized in Table~\ref{table:case_results}.
It can be readily observed that the proposed BS-MRMR model consistently outperforms the frequentist and Bayesian separate models, and HGT model for all types of responses.
The BS-MRMR performs significantly the best in the prediction of counting runtime performance metric, which may be attributed to the shared information from other correlated metrics.
We then further investigate the estimated precision matrix $\boldsymbol{\hat{\Omega}}$ and the corresponding correlation matrix.
In Figure~\ref{Fig:omega_interval} (a), the median values of $\hat{\bm \Omega}$ and the estimated 95\% credible intervals are visualized in matrix bar plots with error bars.
Note that the median values of $\hat{\boldsymbol{\Omega}}$ are standardized in the range of $[-1,1]$ with diagonal elements to be all ones.
Figure~\ref{Fig:omega_interval} (b) plots the median values of correlation matrix with 95\% credible intervals, which are converted from the estimated precision matrix.
This correlation matrix well aligns with the generation of the runtime performance metrics.
For example, the counting metric $\mathbb{Y}_3$ is correlated with the continuous metric $\mathbb{Y}_1$ (i.e., $corr(\xi_1,\xi_3)=0.218$),
since the number of sub-steps that can be executed in five seconds highly depends on the CPU utilization.
As another example, $\mathbb{Y}_2$ is highly correlated with $\mathbb{Y}_4$ (i.e., $corr(\xi_2, \xi_4)=0.637$), since $\mathbb{Y}_4$ is generated by comparing $\mathbb{Y}_2$ with a certain threshold.
Besides, Figure~\ref{Fig:omega_interval} presents the sparsity of the estimated precision matrix $\hat{\bm \Omega}$ with narrow credible intervals,
which demonstrates the effectiveness of the slack-and-slab prior imposed in the precision matrix $\bm \Omega$.

\begin{table}[H]
\begin{center}
\caption{The averages and standard errors (in parenthesis) of loss measures for the real case study in Fog manufacturing.}
\label{table:case_results}
\begin{tabular}{cccccccc}
\hline \hline
&    &$RMSE(N)$  &$RMSE(P)$ &$ME$   \\
\hline
&FS-GLM &1.345 (0.054) &102.9 (45.57)       &0.058 (0.024) \\
&BS-GLM &6.074 (1.344) &338.0 (52.65)       &0.240 (0.055)  \\
&HGT    &3.186 (0.366)  &20.79 (5.723)     & 0.263 (0.023)\\
&BS-MRMR   &0.521 (0.044) &10.24 (0.434)       &0.039 (0.012) \\
\hline \hline
\end{tabular}
\end{center}
\end{table}

\begin{figure}
\begin{center}
\scalebox{0.7}[0.7]{\includegraphics{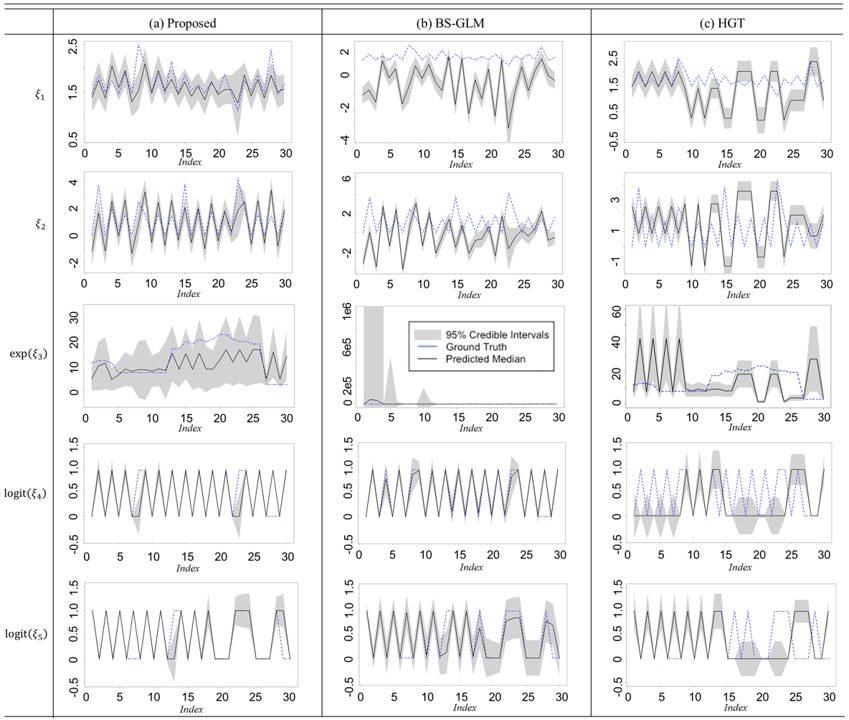}}
\caption{The median of the predicted latent responses $\xi_1, \cdots, \xi_5$ and the associated 95\% credible intervals on the testing data: (a) the proposed BS-MRMR, (b) the BS-GLM, (c) the HGT.}\label{Fig:conf_int}
\end{center}
\end{figure}

Moreover, we also investigate statistical inferences for the uncertainty quantification of the predicted mixed metrics.
Figure~\ref{Fig:conf_int}(a) reports the median of the predicted latent responses $\xi_1, \cdots, \xi_5$ by the BS-MRMR and the associated 95\% credible intervals on the testing data.
In Figure~\ref{Fig:conf_int}, 95\% credible intervals are presented by the shaded region, the median values of the predicted latent variables are plotted in black solid lines, and the true responses are plotted in blue dotted lines.
Figure~\ref{Fig:conf_int}(b) and (c) report the prediction and the associated 95\% credible intervals on the testing data from the BS-GLM and HGT methods respectively.
It is noted that the true responses are mainly contained by the 95\% credible intervals from the proposed BS-MRMR model,
while the 95\% credible intervals of the BS-GLM and HGT do not perform well to cover the true responses.
The narrow credible intervals of the BS-MRMR model can be attributed to the joint modeling of mixed responses and the quantification of hidden associations among these mixed responses.
Note that the narrow credible intervals indicate low uncertainty in predicting runtime performance metrics.
In addition, it is seen that the BS-GLM with Poisson response provides unstable prediction, which leads to extremely large credible intervals (see $\exp{(\boldsymbol{\xi}_3)}$ in Figure~\ref{Fig:conf_int}(b)).
Besides, FS-GLM cannot provide uncertainty quantification, hence its intervals are not available.

\begin{table}[h]
\centering
\caption{Settings for sensitivity study with five factors and two levels at each factor.}
\label{tab:sen_setting}
\begin{tabular}{ccc}
\hline \hline
\multicolumn{1}{c}{\textbf{Factors}} & \multicolumn{1}{c}{\textbf{Level 1}} & \multicolumn{1}{c}{\textbf{Level 2}} \\ \hline
$(a_1, a_2)$ & $(1, 1)$ & $(2,2)$ \\
$(a_3, a_4)$ & $(2p, p)$ & $(p, p)$ \\
$(a_5, a_6)$ & $(q, q(q-1)/2)$ & $(q, q)$ \\
$(\alpha, \lambda)$ & $(q/2, q)$ & $(q, q)$ \\
$(\sigma_0, \sigma_1)$ & $(0.1, 3)$ & $(0.2, 2)$ \\ \hline \hline
\end{tabular}
\end{table}

In addition, we also conduct the sensitivity analysis on the choice of priors with respect to hyper-parameters in the proposed BS-MRMR model.
Specifically, we set a full factorial design for five pairs of hyperparameters as five factors with two levels for each factor. The five factors and levels are listed in Table~\ref{tab:sen_setting}.
The 10-fold cross validations are used to check the prediction performance of the proposed method under $2^5(=32)$ setting of hyper-parameters.
Detailed settings for 32 designs are summarized in Table 8 of the Supplementary Materials.
From the results, one can see there is not significant differences under different setting of prior hyper-parameters, indicating that the proposed BS-MRMR method is not sensitive to the choice of priors.

\begin{table}[h]
\centering
\caption{Sensitivity study results to explore different combinations of priors. See detailed settings of each run in Table 8 of the Supplementary Materials.}
\label{tab:sensitivity}
\resizebox{\textwidth}{!}{%
\begin{tabular}{llll|llll}
\hline \hline
\multicolumn{1}{c}{\textbf{DOE}} & \multicolumn{1}{c}{\textbf{RMSE Normal}} & \multicolumn{1}{c}{\textbf{RMSE Poisson}} & \multicolumn{1}{c|}{\textbf{ME}} & \multicolumn{1}{c}{\textbf{DOE}} & \multicolumn{1}{c}{\textbf{RMSE Normal}} & \multicolumn{1}{c}{\textbf{RMSE Poisson}} & \multicolumn{1}{c}{\textbf{ME}} \\ \hline
1 & 0.515 (0.043) & 10.281 (0.417) & 0.051 (0.010) & 17 & 0.517 (0.044) & 10.248 (0.459) & 0.053 (0.011) \\
2 & 0.516 (0.044) & 10.141 (0.418) & 0.055 (0.011) & 18 & 0.510 (0.043) & 10.234 (0.418) & 0.054 (0.011) \\
3 & 0.520 (0.044) & 10.325 (0.487) & 0.055 (0.011) & 19 & 0.521 (0.044) & 10.385 (0.447) & 0.051 (0.010) \\
4 & 0.519 (0.044) & 10.227 (0.406) & 0.056 (0.011) & 20 & 0.519 (0.043) & 10.490 (0.494) & 0.052 (0.011) \\
5 & 0.514 (0.044) & 10.341 (0.523) & 0.050 (0.010) & 21 & 0.525 (0.044) & 10.239 (0.437) & 0.053 (0.011) \\
6 & 0.515 (0.044) & 10.291 (0.445) & 0.051 (0.010) & 22 & 0.517 (0.044) & 10.224 (0.442) & 0.055 (0.011) \\
7 & 0.522 (0.045) & 10.303 (0.444) & 0.055 (0.010) & 23 & 0.518 (0.044) & 10.406 (0.441) & 0.053 (0.011) \\
8 & 0.524 (0.044) & 10.367 (0.483) & 0.055 (0.012) & 24 & 0.522 (0.044) & 10.170 (0.447) & 0.052 (0.010) \\
9 & 0.517 (0.044) & 10.289 (0.422) & 0.054 (0.010) & 25 & 0.529 (0.044) & 10.318 (0.424) & 0.055 (0.010) \\
10 & 0.515 (0.044) & 10.225 (0.459) & 0.053 (0.011) & 26 & 0.519 (0.044) & 10.185 (0.434) & 0.055 (0.010) \\
11 & 0.517 (0.044) & 10.285 (0.423) & 0.054 (0.010) & 27 & 0.517 (0.044) & 10.225 (0.418) & 0.051 (0.010) \\
12 & 0.515 (0.044) & 10.218 (0.419) & 0.051 (0.010) & 28 & 0.518 (0.044) & 10.366 (0.419) & 0.052 (0.010) \\
13 & 0.518 (0.044) & 10.284 (0.443) & 0.053 (0.011) & 29 & 0.520 (0.043) & 10.310 (0.424) & 0.052 (0.011) \\
14 & 0.513 (0.043) & 10.405 (0.485) & 0.053 (0.011) & 30 & 0.515 (0.043) & 10.386 (0.483) & 0.052 (0.011) \\
15 & 0.525 (0.044) & 10.521 (0.485) & 0.051 (0.010) & 31 & 0.515 (0.044) & 10.285 (0.489) & 0.053 (0.011) \\
16 & 0.521 (0.045) & 10.193 (0.418) & 0.054 (0.012) & 32 & 0.515 (0.044) & 10.359 (0.406) & 0.056 (0.013) \\ \hline \hline
\end{tabular}%
}
\end{table}

To support predictive offloading in Fog manufacturing, the offloading method should determine both the offloading strategies (e.g., randomly offloading, closest distance-based offloading, etc.) and the offloading decisions by considering not only the predicted runtime performance metrics, but also the uncertainty associated with the predictions.
For example, it is confident for the predictive offloading strategy to optimize the offloading decision based on the accurate prediction with low prediction uncertainty.
Thus, the offloading decisions can be optimized via the algorithm in \cite{chen2018predictive} or \cite{zhang2017mobile} based on the predicted runtime performance metrics.
While the high prediction uncertainty will prevent the adoption of the predictive offloading strategy, which highly depends on the accuracy of the predictions.
Hence, other offloading strategies are preferred under this circumstance.

\section{Discussion} \label{sec:conclusion}

This work develops a Bayesian regression for jointly modeling mixed multi-responses to achieve accurate prediction and uncertainty quantification with meaningful model interpretation.
The proposed BS-MRMR method can quantify the hidden associations among mixed responses to improve the prediction performance.
As evidenced in the case study of Fog manufacturing,
the superior prediction performance of the BS-MRMR model with the capability of uncertainty quantification demonstrates its merits to support predictive offloading in Fog computing network.
Not restricted to Fog manufacturing, the proposed method can also be applied in other areas such as health care and material science.
%

There are several directions for future researches. One direction is to investigate how to incorporate the quantified predicted uncertainty in the predictive offloading method by formulating the offloading problem as a chance-constrained optimization problem.
Then the optimized offloading decisions can be more trustworthy to the performance of the predictive models.
Besides predictive offloading, the proposed BS-MRMR model also facilitates the optimization of the Fog computing architecture by evaluating different designs based on the predicted performance metrics.
Another direction is to extend the proposed BS-MRMR model to other types of responses such as censored outcomes and functional responses \citep{sun2017functional}.
For example, one may consider the proposed method for functional mixed responses, which considers runtime performance metrics as time series, hence providing more informative prediction.
Moreover, when the data are functional with no predictors available, several statistical techniques such as spline and wavelet can be applied to create a set of predictors.
Then the proposed BS-MRMR method could accommodate such situations where the priors for group and individual sparsity need to be modified accordingly.
Finally, we note that there are few theoretical results for the model of mixed multivariate responses due to the complex structure in the responses.
For the case of single binary response and single continuous response, \cite{kurum2016time} characterized the binary response using the latent variable and established the asymptotic normality of their estimator.
It will be an interesting to borrow their ideas to investigate the posterior consistency for the proposed BS-MRMR, which is a Bayesian approach of using latent variables.
Note that the proposed method adopt the spike-and-slab priors to enable variable selection,
additional techniques such as Bayes factor are needed to investigate the estimation and selection consistency.

\section{Supplementary Materials}
The supplementary materials for this article contain the following:
(1) detailed derivation of full-conditional distributions;
(2) detailed performance comparison in numerical study;
and (3) detailed full factorial design for sensitivity study of prior settings.

\bibliographystyle{apalike}
\bibliography{reference}

\appendix
\begin{appendices}

\section{Detailed Derivation of Full-Conditional Distributions}

\subsection{Deriving $p(\Tilde{\bm{B}}, \bm{\Omega}, \bm{\tau}, \pi_1, \pi_2, \pi_3, \sigma_{\tau}^2 |\bm{\xi})$}
\begin{align*}
 & p(\Tilde{\bm{B}}, \bm{\Omega}, \bm{\tau}, \pi_1, \pi_2, \pi_3, \sigma_{\tau}^2 |\bm{\xi}) \\
    \propto &p(\sigma_{\tau}^2) p(\pi_1) p(\pi_2) p(\pi_3) p(\Tilde{\bm{B}} | \bm{\Omega}, \pi_1) p(\bm{\tau}|\pi_2,\sigma_{\tau}^2) p(\bm{\Omega} | \pi_3)
    \prod_{i=1}^{n}{p(\bm{\xi}_i|\bm{X},\Tilde{\bm{B}},\bm{\Omega},\bm{\tau})}.\\
    \propto & {|\bm{\Omega}|}^{\frac{n}{2}} \exp \left\{-\frac{1}{2}\tr \left[ \bm{\Omega} (\bm{\Xi}-\sum_{g=1}^G \mathbb{X}_g\bm V_g\Tilde{\bm{B}}_g)^T (\bm{\Xi}-\sum_{g=1}^G \mathbb{X}_g\bm V_g\Tilde{\bm{B}}_g) \right] \right\}\\
    \times & \left\{ \prod_{g=1}^G (1-\pi_1) (2\pi)^{-\frac{p_g q}{2}} |\bm \Omega|^{\frac{p_g}{2}} \exp \left[ -\frac{1}{2} \tr \left( \bm \Omega \Tilde{\bm{B}}_g^T \Tilde{\bm{B}}_g \right) \right] \bm I(\Tilde{\bm{B}}_g \neq \bm{0}) + \pi_1\delta_0(\vec (\Tilde{\bm{B}}_g^T))  \right\} \\
    \times & \prod_{g=1}^G \prod_{j=1}^{p_g} \left\{(1-\pi_2)2(2\pi \sigma_{\tau}^2)^{-\frac{1}{2}} \exp \left[ - \frac{\tau_{g,j}^2}{2\sigma_{\tau}^2} \right] \bm I(\tau_{g,j}>0) + \pi_2\delta_0(\tau_{g,j}) \right\}\\
    \times & \prod_{i<j} \left \{ (1-\pi_3) \exp(-\frac{\omega_{ij}^2}{2 \sigma_0^2}) \bm I(\omega_{ij} = 0) + \pi_3 \exp(-\frac{\omega_{ij}^2}{2 \sigma_1^2}) \bm I(\omega_{ij} \neq 0) \right \} \prod_{i}^p \exp(-\frac{\lambda}{2} \omega_{ii}) \\
    \times & \pi_1^{a_1-1}(1-\pi_1)^{a_2-1} \times \pi_2^{a_3-1}(1-\pi_2)^{a_4-1} \times \pi_3^{a_5-1}(1-\pi_3)^{a_6-1} \times (\sigma_{\tau}^2)^{-2} \exp(-\frac{d}{\sigma_{\tau}^2}).
\end{align*}

\subsection{Deriving $\pi_{B_g}$}

The full-conditional distribution of $\Tilde{\bm{B}_g}$ is
\begin{align}\label{post:B}
\vec{(\Tilde{\boldsymbol{B}}^T_g | \mbox{rest} }) \sim
    (1-\pi_{B_g}) \mathcal{N}_{p_gq}(\vec(\bm M_g^T), \bm \Psi_{p_g}\otimes\boldsymbol{\Sigma}) +
    \pi_{B_g} \delta_0(\vec{(\Tilde{\boldsymbol{B}}^T_g)})
\end{align}
for $g=1,\dots,G$, where $\bm \Psi_{p_g} = (\bm I_{p_g} + \bm V_g \mathbb{X}_g^T \mathbb{X}_g \bm V_g)^{-1}$, $\bm M_g = \bm \Psi_{p_g} \bm V_g \mathbb{X}_g^T (\bm{\Xi}-\sum_{k \neq g}^G \mathbb{X}_k \bm V_k \Tilde{\bm{B}}_k)$, and
\begin{align*}
\pi_{B_g} 
&= \frac{p(\Tilde{\bm{B}}_g = 0 | \mbox{rest} )}{p(\Tilde{\bm{B}}_g = 0 | \mbox{rest} ) + p(\Tilde{\bm{B}}_g \neq 0 | \mbox{rest} )},
\end{align*}
where
\begin{align*}
p(\Tilde{\bm{B}}_g = 0 | \mbox{rest} ) &= \pi_1 \exp \left\{-\frac{1}{2}\tr \left[ \bm{\Omega} (\bm{\Xi}-\sum_{k \neq g}^G \mathbb{X}_k \bm V_k \Tilde{\bm{B}}_k)^T
(\bm{\Xi}-\sum_{k \neq g}^G \mathbb{X}_k \bm V_k \Tilde{\bm{B}}_k) \right] \right\} \\
p(\Tilde{\bm{B}}_g \neq 0 | \mbox{rest} ) &= (1-\pi_1) \exp \left\{-\frac{1}{2}\tr \left[ \bm{\Omega} (\bm{\Xi}-\sum_{g=1}^G \mathbb{X}_g\bm V_g\Tilde{\bm{B}}_g)^T (\bm{\Xi}-\sum_{g=1}^G \mathbb{X}_g\bm V_g\Tilde{\bm{B}}_g) \right] \right\} \\
&\times (2\pi)^{-\frac{p_g q}{2}} |\bm \Omega|^{\frac{p_g}{2}} \exp \left[ -\frac{1}{2} \tr \left( \bm \Omega \Tilde{\bm{B}}_g^T \Tilde{\bm{B}}_g \right) \right] \\
&= (1-\pi_1) \exp \left\{-\frac{1}{2}\tr \left[ \bm{\Omega} (\bm{\Xi}-\sum_{k \neq g}^G \mathbb{X}_k \bm V_k \Tilde{\bm{B}}_k)^T
(\bm{\Xi}-\sum_{k \neq g}^G \mathbb{X}_k \bm V_k \Tilde{\bm{B}}_k) \right] \right\} \\
&\times |\bm \Psi_{p_g}|^{\frac{q}{2}} \exp \left\{ \frac{1}{2}\tr (
\bm{\Omega} \bm{M}_g^T \bm{\Psi}_{p_g}^{-1} \bm{M}_g)
\right\}.
\end{align*}
Therefore,
\begin{align*}
\pi_{B_g} = \frac{\pi_1}{\pi_1 + (1- \pi_1)|\bm \Psi_{p_g}|^{\frac{q}{2}} \exp \left\{ \frac{1}{2}\tr ( \bm{\Omega} \bm{M}_g^T \bm{\Psi}_{p_g}^{-1} \bm{M}_g )
\right\}
}.
\end{align*}

\subsection{Deriving $f(\bm \xi_i \mid \bm y_i, \bm B, \bm \Sigma)$}
\begin{align*}\label{eq:xi}
f(\bm \xi_i \mid \bm y_i, \bm B, \bm \Sigma)
\propto& f(\bm \xi_i \mid \bm B, \bm \Sigma) f(\bm y_i \mid \bm \xi_i, \bm B, \bm \Sigma)   \nonumber \\
\propto& f(\bm \xi_i \mid \bm B, \bm \Sigma) \prod_{j=1}^{l}f(U^{(j)}=u_i^{(j)} \mid \bm \xi_i) \prod_{j=1}^{m}f(Z^{(j)}=z_i^{(j)} \mid \bm \xi_i) \prod_{j=1}^{k}f(W^{(j)}=w_i^{(j)} \mid \bm \xi_i)  \nonumber \\
\propto& |\bm \Omega|^{\frac{1}{2}} \exp \left( -\frac{1}{2}[\bm{\xi}_i - \bm{B}^{T} \bm{x}_i]^T \bm \Omega [\bm{\xi}_i - \bm{B}^{T} \bm{x}_i] \right) \prod_{j=1}^{l} \frac{1}{\sqrt{2 \pi} \sigma^{(j)}} \exp{\{-\frac{(u_i^{(j)} - \mu^{(j)})^2}{2 {\sigma^{(j)}}^{2}}\}}  \nonumber \\
& \prod_{j=1}^{m} \frac{(\lambda^{(j)})^{z_i^{(j)}} \exp{(-\lambda^{(j)})}}{z_i^{(j)} !} \cdot \prod_{j=1}^{k} (\gamma^{(j)})^{w_i^{(j)}} (1 - \gamma^{(j)})^{1 - w_i^{(j)}},
\end{align*}

\section{Detailed Performance Comparison in Numerical Study}

In this section, we summarized the performance comparison results for numerical study.
Specifically, the parameter estimation errors are summarized in Table~\ref{table:sim1}; the variable selection errors are summarized in Table~\ref{table:simfsl}; the prediction errors of $p=20$ and $p=80$ are summarized in Tables~\ref{table:p20} and \ref{table:p80}, respectively.

\begin{table}[H]
\scriptsize
\begin{center}
\caption{The averages and standard errors (in parenthesis) of parameter estimation errors of $\hat{\boldsymbol{B}}$ and $\hat{\boldsymbol{\Omega}}$
for the comparison of BS-MRMR (proposed), FS-GLM and BS-GLM.}
\label{table:sim1}
\begin{tabular}{ccccccccccccccc}
\hline \hline
&&     &$\bm \Omega_1$  &$\bm \Omega_2$ &$\bm \Omega_3$ &$\bm \Omega_4$ &$\bm \Omega_5$  \\
\cline{1-8}
&&&&&$L(\hat{\bm B})$\\
\hline
\multirow {6}*{$p=20$}
&\multirow {3}*{$\bm B_1$}
&FS-GLM  &0.460 (0.277) &0.322 (0.145) &0.388 (0.204) &0.547 (0.289) &0.632 (0.346)  \\
&&BS-GLM &0.253 (0.007) &0.249 (0.007) &0.254 (0.007) &0.251 (0.007) &0.273 (0.007)   \\
&&BS-MRMR &0.148 (0.005) &0.151 (0.005) &0.143 (0.006) &0.147 (0.005) &0.161 (0.007)  \\
\cline{3-8}
&\multirow {3}*{$\bm B_2$}
&FS-GLM  &0.585 (0.304) &0.178 (0.004) &0.308 (0.112) &0.667 (0.246) &0.470 (0.191)  \\
&&BS-GLM &0.244 (0.006) &0.241 (0.005) &0.255 (0.007) &0.249 (0.007) &0.278 (0.008)  \\
&&BS-MRMR &0.141 (0.004) &0.138 (0.005) &0.149 (0.005) &0.154 (0.006) &0.156 (0.006)  \\
\midrule
\multirow {6}*{$p=80$}
&\multirow {3}*{$\bm B_1$}
&FS-GLM  &0.130 (0.006) &0.119 (0.001) &0.122 (0.003) &0.109 (0.001) &0.104 (0.001)  \\
&&BS-GLM &0.226 (0.002) &0.225 (0.002) &0.223 (0.002) &0.202 (0.002) &0.194 (0.002)  \\
&&BS-MRMR  &0.081 (0.001) &0.083 (0.001) &0.082 (0.001) &0.078 (0.001) &0.083 (0.001)  \\
\cline{3-8}
&\multirow {3}*{$\bm B_2$}
&FS-GLM  &0.122 (0.013) &0.102 (0.001) &0.101 (0.001) &0.091 (0.001) &0.086 (0.000)  \\
&&BS-GLM &0.225 (0.003) &0.221 (0.002) &0.220 (0.003) &0.201 (0.002) &0.189 (0.002)  \\
&&BS-MRMR  &0.063 (0.001) &0.062 (0.001) &0.063 (0.001) &0.060 (0.001) &0.062 (0.001)   \\
\midrule
&&&&&$L(\hat{\bm \Omega})$\\
\hline
\multirow {4}*{$p=20$}
&\multirow {2}*{$\bm B_1$}
&BS-GLM &2.963 (0.171) &3.194 (0.201) &3.600 (0.380) &2.998 (0.199) &4.387 (0.685)  \\
&&BS-MRMR  & 0.270 (0.004) &0.282 (0.005) &0.271 (0.004) &0.235 (0.005) &0.247 (0.004)  \\
\cline{3-8}
&\multirow {2}*{$\bm B_2$}
&BS-GLM &3.363 (0.276) &3.447 (0.288) &3.163 (0.285) &3.063 (0.223) &3.267 (0.293)  \\
&&BS-MRMR  & 0.271 (0.003) &0.290 (0.004) &0.276 (0.005) &0.229 (0.005) &0.255 (0.005) \\
\midrule
\multirow {4}*{$p=80$}
&\multirow {2}*{$\bm B_1$}
&BS-GLM &1.238 (0.267) &0.801 (0.123) &1.081 (0.228) &1.004 (0.183) &0.500 (0.073)  \\
&&BS-MRMR  & 0.196 (0.002) &0.197 (0.001) &0.200 (0.001) &0.173 (0.002) &0.168 (0.002)  \\
\cline{3-8}
&\multirow {2}*{$\bm B_2$}
&BS-GLM &0.865 (0.208) &0.728 (0.119) &0.897 (0.152) &0.563 (0.087) &0.800 (0.155)  \\
&&BS-MRMR  & 0.188 (0.001) &0.190 (0.001) &0.193 (0.001) &0.164 (0.001) &0.169 (0.002)  \\
\hline
\hline
\end{tabular}
\end{center}
\end{table}

\begin{table}[H]
\scriptsize
\begin{center}
\caption{The averages and standard errors (in parenthesis) of variable selection errors in $\hat{\boldsymbol{B}}$ and $\hat{\boldsymbol{\Omega}}$ for the comparison of BS-MRMR (proposed), FS-GLM, BS-GLM, and HGT.}
\label{table:simfsl}
\begin{tabular}{ccccccccccccccc}
\hline \hline
&&     &$\bm \Omega_1$  &$\bm \Omega_2$ &$\bm \Omega_3$ &$\bm \Omega_4$ &$\bm \Omega_5$  \\
\cline{1-8}
&&&&&$FSL\hat{(\bm {B})}$\\
\hline
\multirow {8}*{$p=20$}
&\multirow {4}*{$\bm B_1$}
&FS-GLM  &0.327 (0.006) &0.332 (0.007) &0.334 (0.006) &0.308 (0.006) &0.320 (0.007) \\
&&BS-GLM &0.240 (0.005) &0.233 (0.005) &0.245 (0.005) &0.221 (0.005) &0.228 (0.005)  \\
&&HGT &0.034 (0.005)	&0.025 (0.005)	&0.021 (0.005)	&0.041 (0.005)	&0.032 (0.005) \\
&&BS-MRMR  &0.037 (0.005) &0.028 (0.005) &0.030 (0.005) &0.022 (0.004) &0.019 (0.004)  \\
\cline{3-8}
&\multirow {4}*{$\bm B_2$}
&FS-GLM  &0.327 (0.006) &0.326 (0.005) &0.337 (0.006) &0.316 (0.007) &0.333 (0.008) \\
&&BS-GLM &0.240 (0.004) &0.244 (0.006) &0.240 (0.006) &0.222 (0.005) &0.230 (0.005) \\
&&HGT &0.026 (0.003)	&0.078 (0.007)	&0.037 (0.005)	&0.034 (0.005)	&0.023 (0.005) \\
&&BS-MRMR  &0.025 (0.004) &0.023 (0.004) &0.029 (0.005) &0.028 (0.006) &0.021 (0.004)  \\
\midrule
\multirow {8}*{$p=80$}
&\multirow {4}*{$\bm B_1$}
&FS-GLM  &0.276 (0.003) &0.271 (0.003) &0.275 (0.003) &0.243 (0.002) &0.228 (0.002)  \\
&&BS-GLM &0.207 (0.002) &0.206 (0.001) &0.203 (0.001) &0.193 (0.001) &0.186 (0.001) \\
&&HGT &0.045 (0.005)	&0.039 (0.004)	&0.036 (0.005)	&0.035 (0.005)	&0.026 (0.003) \\
&&BS-MRMR  &0.031 (0.002) &0.030 (0.003) &0.031 (0.002) &0.029 (0.003) &0.027 (0.002) \\
\cline{3-8}
&\multirow {4}*{$\bm B_2$}
&FS-GLM  &0.232 (0.003) &0.225 (0.003) &0.229 (0.002) &0.195 (0.003) &0.171 (0.003) \\
&&BS-GLM &0.160 (0.002) &0.158 (0.002) &0.156 (0.001) &0.148 (0.001) &0.136 (0.001)  \\
&&HGT  &0.028 (0.002)	&0.021 (0.002)	&0.032 (0.003)	&0.023 (0.002)	&0.019 (0.002) \\
&&BS-MRMR  &0.010 (0.002) &0.008 (0.001) &0.010 (0.001) &0.006 (0.001) &0.005 (0.001) \\
\midrule
&&&&&$FSL\hat{(\bm {\Omega})}$\\
\hline
\multirow {4}*{$p=20$}
&\multirow {2}*{$\bm B_1$}
&BS-GLM &0.500 (0.000) &0.500 (0.000) &0.667 (0.000) &0.333 (0.000) &0.333 (0.000)  \\
&&BS-MRMR  &0.361 (0.012) &0.352 (0.011) &0.322 (0.013) &0.373 (0.016) &0.359 (0.017) \\
\cline{3-8}
&\multirow {2}*{$\bm B_2$}
&BS-GLM  &0.500 (0.000) &0.500 (0.000) &0.667 (0.000) &0.333 (0.000) &0.333 (0.000) \\
&&BS-MRMR  &0.363 (0.012) &0.374 (0.014) &0.331 (0.012) &0.362 (0.014) &0.393 (0.016)  \\
\midrule
\multirow {4}*{$p=80$}
&\multirow {2}*{$\bm B_1$}
&BS-GLM &0.240 (0.000) &0.240 (0.000) &0.347 (0.000) &0.133 (0.000) &0.053 (0.000)  \\
&&BS-MRMR  &0.359 (0.006) &0.360 (0.005) &0.395 (0.006) &0.301 (0.006) &0.279 (0.006) \\
\cline{3-8}
&\multirow {2}*{$\bm B_2$}
&BS-GLM &0.240 (0.000) &0.240 (0.000) &0.347 (0.000) &0.133 (0.000) &0.053 (0.000) \\
&&BS-MRMR  &0.356 (0.005) &0.346 (0.005) &0.385 (0.007) &0.291 (0.006) &0.277 (0.006) \\
\hline \hline
\end{tabular}
\end{center}
\end{table}

\begin{table}[H]
\scriptsize
\begin{center}
\caption{The averages and standard errors (in parenthesis) of prediction errors under $p=20$ for the comparison of BS-MRMR (proposed), FS-GLM, BS-GLM, and HGT.}
\label{table:p20}
\begin{tabular}{ccccccccccccccc}
\hline \hline
&&     &$\bm \Omega_1$  &$\bm \Omega_2$ &$\bm \Omega_3$ &$\bm \Omega_4$ &$\bm \Omega_5$  \\
\hline
\multirow {8}*{$RMSE(N)$}
&\multirow {4}*{$\bm B_1$}
&FS-GLM  &1.319 (0.009) &1.360 (0.012) &1.313 (0.013) &1.221 (0.011) &1.364 (0.011) \\
&&BS-GLM &1.312 (0.011) &1.348 (0.012) &1.308 (0.012) &1.201 (0.010) &1.343 (0.011) \\
&&HGT  &1.301 (0.012)	&1.417 (0.012)	&1.428 (0.013)	&1.315 (0.011)	&1.313 (0.011) \\
&&BS-MRMR  &1.265 (0.010) &1.297 (0.010) &1.263 (0.010) &1.167 (0.009) &1.310 (0.010)  \\
\cline{3-8}
&\multirow {4}*{$\bm B_2$}
&FS-GLM  &1.323 (0.011) &1.359 (0.013) &1.334 (0.012) &1.247 (0.011) &1.350 (0.012) \\
&&BS-GLM &1.316 (0.012) &1.357 (0.014) &1.315 (0.012) &1.236 (0.013) &1.349 (0.011) \\
&&HGT  &1.523 (0.015)	&1.412 (0.013)	&1.421 (0.012)	&1.282 (0.010)	&1.302 (0.011) \\
&&BS-MRMR  &1.273 (0.011) &1.308 (0.011) &1.285 (0.011) &1.190 (0.010) &1.299 (0.011) \\
\midrule
\multirow {8}*{$RMSE(P)$}
&\multirow {4}*{$\bm B_1$}
&FS-GLM  &19.91 (2.553) &31.91 (9.315) &31.85 (7.867) &14.59 (1.053) &19.34 (2.350)  \\
&&BS-GLM &29.85 (4.932) &48.06 (13.43) &57.34 (17.21) &17.50 (1.475) &33.70 (6.867)  \\
&&HGT  &15.55 (2.872)	&13.02 (2.552)	&14.77 (2.686)	&15.53 (3.012)	&16.48 (3.421)  \\
&&BS-MRMR  &11.05 (0.492) &11.59 (0.764) &11.90 (0.586) &10.07 (0.528) &11.07 (0.623)  \\
\cline{3-8}
&\multirow {4}*{$\bm B_2$}
&FS-GLM  &22.52 (5.256) &19.33 (3.710) &19.09 (2.146) &19.09 (2.687) &20.03 (2.225)  \\
&&BS-GLM &36.75 (11.15) &27.33 (4.860) &34.56 (8.600) &25.54 (4.575) &26.51 (3.284)  \\
&&HGT  &18.56 (3.235)	&17.24 (3.642)	&18.35 (3.854)	&17.19 (3.256)	&16.49 (2.985)  \\
&&BS-MRMR  &14.36 (2.683) &10.26 (0.486) &11.47 (0.492) &10.18 (0.711) &11.88 (0.728) \\
\midrule
\multirow {8}*{$ME$}
&\multirow {4}*{$\bm B_1$}
&FS-GLM  &0.301 (0.008) &0.294 (0.009) &0.326 (0.011) &0.294 (0.011) &0.259 (0.009) \\
&&BS-GLM &0.276 (0.005) &0.271 (0.005) &0.294 (0.008) &0.262 (0.007) &0.237 (0.007) \\
&&HGT &0.275 (0.006)	&0.352 (0.006)	&0.255 (0.006)	&0.315 (0.006)	&0.210 (0.006) \\
&&BS-MRMR  &0.263 (0.006) &0.259 (0.006) &0.276 (0.006) &0.255 (0.006) &0.221 (0.005) \\
\cline{3-8}
&\multirow {4}*{$\bm B_2$}
&FS-GLM  &0.307 (0.010) &0.315 (0.009) &0.305 (0.008) &0.277 (0.008) &0.258 (0.009) \\
&&BS-GLM &0.270 (0.007) &0.278 (0.007) &0.277 (0.006) &0.254 (0.005) &0.240 (0.005) \\
&&HGT &0.276 (0.007)	&0.292 (0.007)	&0.289 (0.007)	&0.254 (0.007)	&0.192 (0.004) \\
&&BS-MRMR  &0.257 (0.007) &0.266 (0.007) &0.263 (0.005) &0.240 (0.005) &0.231 (0.005) \\
\hline \hline
\end{tabular}
\end{center}
\end{table}

\begin{table}[H]
\scriptsize
\begin{center}
\caption{The averages and standard errors (in parenthesis) of prediction errors under $p=80$ for  the comparison of BS-MRMR (proposed), FS-GLM, BS-GLM, and HGT.}
\label{table:p80}
\begin{tabular}{ccccccccccccccc}
\hline \hline
&&     &$\bm \Omega_1$  &$\bm \Omega_2$ &$\bm \Omega_3$ &$\bm \Omega_4$ &$\bm \Omega_5$  \\
\hline
\multirow {8}*{$RMSE(N)$}
&\multirow {4}*{$\bm B_1$}
&FS-GLM  &1.511 (0.009) &1.528 (0.008) &1.525 (0.009) &1.382 (0.008) &1.469 (0.009)  \\
&&BS-GLM &2.056 (0.029) &2.058 (0.030) &2.020 (0.028) &1.864 (0.027) &2.014 (0.034)  \\
&&HGT &1.387 (0.010)	&1.402 (0.009)	&1.381 (0.008)	&1.293 (0.008)	&1.365 (0.009) \\
&&BS-MRMR  &1.379 (0.009) &1.382 (0.008) &1.380 (0.008) &1.232 (0.007) &1.311 (0.006) \\
\cline{3-8}
&\multirow {4}*{$\bm B_2$}
&FS-GLM  &1.468 (0.008) &1.465 (0.008) &1.440 (0.007) &1.327 (0.007) &1.409 (0.009) \\
&&BS-GLM &1.937 (0.033) &1.985 (0.032) &1.945 (0.032) &1.826 (0.026) &1.896 (0.027) \\
&&HGT &1.377 (0.009)	&1.427 (0.011)	&1.347 (0.008)	&1.245 (0.005)	&1.295 (0.006)\\
&&BS-MRMR  &1.342 (0.008) &1.346 (0.006) &1.328 (0.007) &1.204 (0.006) &1.273 (0.007)\\
\midrule
\multirow {8}*{$RMSE(P)$}
&\multirow {4}*{$\bm B_1$}
&FS-GLM  &23.45 (2.765) &27.43 (6.140) &19.29 (1.709) &10.42 (0.569) &7.053 (0.322) \\
&&BS-GLM &269.9 (32.45) &230.8 (26.75) &216.6 (33.99) &137.1 (26.53) &49.64 (11.10)   \\
&&HGT &11.58 (2.325)	&11.38 (2.266)	&15.31 (3.643)	&17.32 (2.897)	&10.48 (1.925)\\
&&BS-MRMR  &10.43 (0.290) &9.979 (0.273) &10.61 (0.333) &8.020 (0.290) &6.257 (0.246) \\
\cline{3-8}
&\multirow {4}*{$\bm B_2$}
&FS-GLM  &29.75 (9.125) &27.20 (8.636) &17.31 (2.888) &7.455 (0.453) &5.670 (0.231) \\
&&BS-GLM &238.6 (32.63) &232.1 (33.92) &171.5 (18.69) &105.2 (20.32) &35.92 (7.293)  \\
&&HGT &9.870 (0.464)	&8.970 (0.356)	&9.129 (0.376)	&9.812 (0.472)	&9.823 (0.436)\\
&&BS-MRMR  &8.752 (0.310) &8.210 (0.260) &8.052 (0.297) &5.950 (0.210) &5.154 (0.197) \\
\midrule
\multirow {8}*{$ME$}
&\multirow {4}*{$\bm B_1$}
&FS-GLM  &0.491 (0.004) &0.484 (0.004) &0.490 (0.003) &0.483 (0.004) &0.481 (0.004) \\
&&BS-GLM &0.448 (0.004) &0.451 (0.004) &0.443 (0.004) &0.439 (0.005) &0.419 (0.004)  \\
&&HGT &0.388 (0.004)	&0.344 (0.004)	&0.356 (0.004)	&0.335 (0.003)	&0.362 (0.004)\\
&&BS-MRMR  &0.335 (0.003) &0.344 (0.004) &0.335 (0.004) &0.315 (0.003) &0.296 (0.003)  \\
\cline{3-8}
&\multirow {4}*{$\bm B_2$}
&FS-GLM  &0.480 (0.004) &0.488 (0.003) &0.486 (0.003) &0.481 (0.005) &0.467 (0.004)  \\
&&BS-GLM &0.451 (0.005) &0.446 (0.004) &0.451 (0.004) &0.439 (0.004) &0.421 (0.004) \\
&&HGT &0.378 (0.004)	&0.396 (0.004)	&0.382 (0.004)	&0.341 (0.003)	&0.346 (0.004)\\
&&BS-MRMR  &0.348 (0.004) &0.360 (0.004) &0.349 (0.004) &0.333 (0.004) &0.307 (0.003)  \\
\hline \hline
\end{tabular}
\end{center}
\end{table}

Besides the evaluation for the scenarios when $n>p$, we also consider the scenarios of $n<p$ where $n=50, p=80$. Similar results are summarized in  Table~\ref{table:n50loss} for the parameter estimation errors $L(\hat{\bm B})$, Table~\ref{table:n50fsl} for the variable selection errors $FSL$, and Table~\ref{table:n50errors} for the prediction errors.

\begin{table}[H]
\scriptsize
\begin{center}
\caption{The averages and standard errors (in parenthesis) of parameter estimation errors of $\hat{\boldsymbol{B}}$ and $\hat{\boldsymbol{\Omega}}$
for the comparison of BS-MRMR (proposed), FS-GLM, and BS-GLM when $n=50, p=80$.}
\label{table:n50loss}
\begin{tabular}{ccccccccccccccc}
\hline \hline
&&     &$\bm \Omega_1$  &$\bm \Omega_2$ &$\bm \Omega_3$ &$\bm \Omega_4$ &$\bm \Omega_5$  \\
\cline{1-8}
&&&&&$L(\hat{\bm B})$\\
\hline
&\multirow {3}*{$\bm B_1$}
&FS-GLM  &0.127 (0.001)	&0.127 (0.001)	&0.125 (0.001)	&0.121 (0.001)	&0.118 (0.001) \\
&&BS-GLM &0.155 (0.001)	&0.156 (0.001)	&0.156 (0.001)	&0.147 (0.001)	&0.153 (0.001) \\
&&BS-MRMR &0.107 (0.001)	&0.109 (0.001)	&0.108 (0.001)	&0.103 (0.001)	&0.105 (0.001) \\
\cline{3-8}
&\multirow {3}*{$\bm B_2$}
&FS-GLM  &0.107 (0.001)	&0.107 (0.001)	&0.107 (0.001)	&0.100 (0.001)	&0.098 (0.001) \\
&&BS-GLM &0.139 (0.001)	&0.142 (0.001)	&0.141 (0.001)	&0.132 (0.001)	&0.139 (0.001) \\
&&BS-MRMR &0.082 (0.001)	&0.084 (0.001)	&0.083 (0.001)	&0.080 (0.001)	&0.078 (0.001) \\
\midrule
&&&&&$L(\hat{\bm \Omega})$\\
\hline
&\multirow {1}*{$\bm B_1$}
&BS-MRMR  & 0.205 (0.001)	&0.208 (0.001)	&0.203 (0.001)	&0.173 (0.002)	&0.161 (0.001) \\
\cline{3-8}
&\multirow {1}*{$\bm B_2$}
&BS-MRMR  & 0.200 (0.001)	&0.202 (0.001)	&0.197 (0.001)	&0.169 (0.001)	&0.158 (0.001) \\

\hline
\hline
\end{tabular}
\end{center}
\end{table}

\begin{table}[H]
\scriptsize
\begin{center}
\caption{The averages and standard errors (in parenthesis) of variable selection errors in $\hat{\boldsymbol{B}}$ and $\hat{\boldsymbol{\Omega}}$ for the comparison of BS-MRMR (proposed), FS-GLM, and BS-GLM when $n=50, p=80$.}
\label{table:n50fsl}
\begin{tabular}{ccccccccccccccc}
\hline \hline
&&     &$\bm \Omega_1$  &$\bm \Omega_2$ &$\bm \Omega_3$ &$\bm \Omega_4$ &$\bm \Omega_5$  \\
\cline{1-8}
&&&&&$L(\hat{\bm B})$\\
\hline
&\multirow {3}*{$\bm B_1$}
&FS-GLM  &0.238 (0.001)	&0.238 (0.002)	&0.234 (0.002)	&0.225 (0.001)	&0.214 (0.001)\\
&&BS-GLM &0.396 (0.000)	&0.396 (0.000)	&0.396 (0.000)	&0.396 (0.000)	&0.396 (0.000)\\
&&BS-MRMR &0.078 (0.004)	&0.078 (0.003)	&0.083 (0.003)	&0.072 (0.004)	&0.064 (0.003)\\
\cline{3-8}
&\multirow {3}*{$\bm B_2$}
&FS-GLM  &0.185 (0.002)	&0.185 (0.002)	&0.187 (0.002)	&0.168 (0.002)	&0.162 (0.002)\\
&&BS-GLM &0.375 (0.000)	&0.375 (0.000)	&0.375 (0.000)	&0.375 (0.000)	&0.375 (0.000)\\
&&BS-MRMR &0.042 (0.002)	&0.050 (0.003)	&0.047 (0.002)	&0.042 (0.003)	&0.034 (0.002)\\

\hline
\hline
\end{tabular}
\end{center}
\end{table}

\begin{table}[H]
\scriptsize
\begin{center}
\caption{The averages and standard errors (in parenthesis) of prediction errors for the comparison of BS-MRMR (proposed), FS-GLM, and BS-GLM when $n=50, p=80$.}
\label{table:n50errors}
\begin{tabular}{ccccccccccccccc}
\hline \hline
&&     &$\bm \Omega_1$  &$\bm \Omega_2$ &$\bm \Omega_3$ &$\bm \Omega_4$ &$\bm \Omega_5$  \\
\hline
\multirow {6}*{$RMSE(N)$}
&\multirow {3}*{$\bm B_1$}
&FS-GLM  &1.660 (0.018)	&1.665 (0.014)	&1.641 (0.014)	&1.530 (0.010)	&1.561 (0.010)\\
&&BS-GLM &2.226 (0.025)	&2.228 (0.023)	&2.218 (0.023)	&2.007 (0.021)	&2.112 (0.021)\\
&&BS-MRMR  &1.524 (0.013)	&1.533 (0.014)	&1.521 (0.015)	&1.391 (0.009)	&1.427 (0.009)\\
\cline{3-8}
&\multirow {3}*{$\bm B_2$}
&FS-GLM  &1.551 (0.011)	&1.556 (0.010)	&1.520 (0.010)	&1.406 (0.010)	&1.478 (0.010)\\
&&BS-GLM &2.152 (0.020)	&2.221 (0.024)	&2.169 (0.021)	&1.940 (0.017)	&2.094 (0.023)\\
&&BS-MRMR  &1.435 (0.011)	&1.455 (0.009)	&1.419 (0.011)  &1.290 (0.009)	&1.345 (0.009)\\
\midrule
\multirow {6}*{$RMSE(P)$}
&\multirow {3}*{$\bm B_1$}
&FS-GLM  &14.18 (0.910)	&11.77 (0.555)	&15.23 (3.660)	&9.474 (0.437)	&7.362 (0.385)\\
&&BS-GLM &12.02 (0.490)	&10.66 (0.472)	&10.92 (0.491)	&9.270 (0.428)	&7.527 (0.398)\\
&&BS-MRMR  &11.67 (0.473)	&10.38 (0.472)	&10.61 (0.488)	&8.852 (0.409)	&7.073 (0.384)\\
\cline{3-8}
&\multirow {3}*{$\bm B_2$}
&FS-GLM  &9.336 (0.460)	&9.380 (0.491)	&9.952 (0.761)	&6.411 (0.327)	&5.225 (0.302)\\
&&BS-GLM &8.408 (0.351)	&8.349 (0.339)	&8.416 (0.346)	&6.267 (0.274)	&5.162 (0.281)\\
&&BS-MRMR  &8.159 (0.346)	&8.110 (0.333)	&8.186 (0.346)	&6.018 (0.273)	&4.797 (0.274)\\
\midrule
\multirow {6}*{$ME$}
&\multirow {3}*{$\bm B_1$}
&FS-GLM  &0.486 (0.004)	&0.492 (0.005)	&0.489 (0.004)	&0.502 (0.005)	&0.485 (0.005)\\
&&BS-GLM &0.505 (0.004)	&0.504 (0.005)	&0.496 (0.003)	&0.498 (0.004)	&0.503 (0.005)\\
&&BS-MRMR  &0.388 (0.005)	&0.386 (0.006)	&0.391 (0.005)	&0.370 (0.005)	&0.357 (0.005)\\
\cline{3-8}
&\multirow {3}*{$\bm B_2$}
&FS-GLM  &0.487 (0.005)	&0.491 (0.004)	&0.487 (0.004)	&0.492 (0.004)	&0.488 (0.005)\\
&&BS-GLM &0.492 (0.003)	&0.503 (0.004)	&0.506 (0.004)	&0.499 (0.004)	&0.507 (0.005)\\
&&BS-MRMR  &0.393 (0.005)	&0.392 (0.005)	&0.397 (0.005)	&0.390 (0.006)	&0.352 (0.006)\\
\hline \hline
\end{tabular}
\end{center}
\end{table}

\section{A Full Factorial Design for Sensitivity Study of Prior Settings}

In this section, we summarized the settings for a full factorial design with five factors in two levels (see Table~\ref{tab:sensedoe}).

\begin{table}[H]
\footnotesize
\centering
\caption{A summary of a full factorial design for sensitivity study of prior settings.}
\label{tab:sensedoe}
\begin{tabular}{llllll}
\hline\hline
DOE & $(a_1,a_2)$ & $(a_3,a_4)$ & $(a_5,a_6)$ & $(\alpha,\lambda)$ & $(\sigma_0, \sigma_1)$ \\ \hline
1 & $(1, 1)$ & $(2p, p)$ & $(q, q(q-1)/2)$ & $(q/2, q)$ & $(0.1, 3)$ \\
2 & $(2, 2)$ & $(2p, p)$ & $(q, q(q-1)/2)$ & $(q/2, q)$ & $(0.1, 3)$ \\
3 & $(1, 1)$ & $(p, p)$ & $(q, q(q-1)/2)$ & $(q/2, q)$ & $(0.1, 3)$ \\
4 & $(2, 2)$ & $(p, p)$ & $(q, q(q-1)/2)$ & $(q/2, q)$ & $(0.1, 3)$ \\
5 & $(1, 1)$ & $(2p, p)$ & $(q, q)$ & $(q/2, q)$ & $(0.1, 3)$ \\
6 & $(2, 2)$ & $(2p, p)$ & $(q, q)$ & $(q/2, q)$ & $(0.1, 3)$ \\
7 & $(1, 1)$ & $(p, p)$ & $(q, q)$ & $(q/2, q)$ & $(0.1, 3)$ \\
8 & $(2, 2)$ & $(p, p)$ & $(q, q)$ & $(q/2, q)$ & $(0.1, 3)$ \\
9 & $(1, 1)$ & $(2p, p)$ & $(q, q(q-1)/2)$ & $(q, q)$ & $(0.1, 3)$ \\
10 & $(2, 2)$ & $(2p, p)$ & $(q, q(q-1)/2)$ & $(q, q)$ & $(0.1, 3)$ \\
11 & $(1, 1)$ & $(p, p)$ & $(q, q(q-1)/2)$ & $(q, q)$ & $(0.1, 3)$ \\
12 & $(2, 2)$ & $(p, p)$ & $(q, q(q-1)/2)$ & $(q, q)$ & $(0.1, 3)$ \\
13 & $(1, 1)$ & $(2p, p)$ & $(q, q)$ & $(q, q)$ & $(0.1, 3)$ \\
14 & $(2, 2)$ & $(2p, p)$ & $(q, q)$ & $(q, q)$ & $(0.1, 3)$ \\
15 & $(1, 1)$ & $(p, p)$ & $(q, q)$ & $(q, q)$ & $(0.1, 3)$ \\
16 & $(2, 2)$ & $(p, p)$ & $(q, q)$ & $(q, q)$ & $(0.1, 3)$ \\
17 & $(1, 1)$ & $(2p, p)$ & $(q, q(q-1)/2)$ & $(q/2, q)$ & $(0.2, 2)$ \\
18 & $(2, 2)$ & $(2p, p)$ & $(q, q(q-1)/2)$ & $(q/2, q)$ & $(0.2, 2)$ \\
19 & $(1, 1)$ & $(p, p)$ & $(q, q(q-1)/2)$ & $(q/2, q)$ & $(0.2, 2)$ \\
20 & $(2, 2)$ & $(p, p)$ & $(q, q(q-1)/2)$ & $(q/2, q)$ & $(0.2, 2)$ \\
21 & $(1, 1)$ & $(2p, p)$ & $(q, q)$ & $(q/2, q)$ & $(0.2, 2)$ \\
22 & $(2, 2)$ & $(2p, p)$ & $(q, q)$ & $(q/2, q)$ & $(0.2, 2)$ \\
23 & $(1, 1)$ & $(p, p)$ & $(q, q)$ & $(q/2, q)$ & $(0.2, 2)$ \\
24 & $(2, 2)$ & $(p, p)$ & $(q, q)$ & $(q/2, q)$ & $(0.2, 2)$ \\
25 & $(1, 1)$ & $(2p, p)$ & $(q, q(q-1)/2)$ & $(q, q)$ & $(0.2, 2)$ \\
26 & $(2, 2)$ & $(2p, p)$ & $(q, q(q-1)/2)$ & $(q, q)$ & $(0.2, 2)$ \\
27 & $(1, 1)$ & $(p, p)$ & $(q, q(q-1)/2)$ & $(q, q)$ & $(0.2, 2)$ \\
28 & $(2, 2)$ & $(p, p)$ & $(q, q(q-1)/2)$ & $(q, q)$ & $(0.2, 2)$ \\
29 & $(1, 1)$ & $(2p, p)$ & $(q, q)$ & $(q, q)$ & $(0.2, 2)$ \\
30 & $(2, 2)$ & $(2p, p)$ & $(q, q)$ & $(q, q)$ & $(0.2, 2)$ \\
31 & $(1, 1)$ & $(p, p)$ & $(q, q)$ & $(q, q)$ & $(0.2, 2)$ \\
32 & $(2, 2)$ & $(p, p)$ & $(q, q)$ & $(q, q)$ & $(0.2, 2)$ \\ \hline\hline
\end{tabular}%
\end{table}
\end{appendices}

\end{document}